%% file: ManticetalCSTEv7-arXiv.tex
\documentclass[
12pt]{elsarticle}

\usepackage{amsmath}
\usepackage{amsfonts}
\usepackage{amssymb}
\usepackage{amsmath}
\usepackage{color}
\usepackage{amsmath,amssymb,amsthm,mathrsfs,epsf,a4,color,graphicx,eucal}

\usepackage{graphicx,subfigure}
\usepackage{epsfig}
\usepackage{multirow}
\usepackage[nodots]{numcompress}

\def\espacio{0.2}

\input defsR

\journal{CSTE}

\begin{document}

\begin{frontmatter}


\title{Application of a linear elastic -  brittle interface model to the crack initiation and propagation at fibre-matrix interface under biaxial transverse loads. \tnoteref{t1}}

\author[]{V. Manti\v{c}}
\ead{mantic@us.es}
\author[]{L. T\'avara \corref{cor1}}
\ead{ltavara@us.es}
\author[]{A. Bl\'{a}zquez}
\ead{abg@us.es}
\author[]{E. Graciani}
\ead{egraciani@us.es}
\author[]{F. Par\'is}
\ead{fparis@us.es}

\cortext[cor1]{Corresponding author. Tel.:+34 954487300; Fax: +34 954461637}

\address{Grupo de Elasticidad y Resistencia de Materiales,\\
Escuela T\'{e}cnica Superior de Ingenier\'{\i}a, Universidad de Sevilla,\\
\vspace{-0.1cm}Camino de los Descubrimientos s/n, ES-41092 Sevilla, Spain}

\title{Application of a linear elastic - brittle interface model to the crack initiation and propagation at fibre-matrix interface under biaxial transverse loads}


\author{}
\address{}

\begin{abstract}
{\small
The crack onset and propagation   at the
fibre-matrix interface in a composite  under    tensile/compressive remote biaxial
transverse loads is studied by   a new linear elastic - (perfectly) brittle interface model.
In this model the interface is represented by a continuous distribution of
springs which simulates the presence of a thin elastic layer.
The     constitutive law  for the continuous distribution  of normal and tangential of initially linear elastic springs
  takes into account   possible   frictionless elastic contact between fibre and
matrix once a portion of the interface is broken.
A brittle failure criterion is employed for the distribution of
springs, which enables the study of  crack onset and propagation. This interface failure criterion
takes into account the variation of the interface fracture toughness with the fracture
mode mixity.
The main advantages of the present interface model are its simplicity, robustness and its computational efficiency when   the so-called  sequentially linear analysis  is applied.
Moreover, in the present plane strain problem of a single fibre embedded in a matrix subjected to uniform remote transverse loads, this model can be used to obtain analytic predictions of interface crack onset.
The numerical results provided by a 2D boundary element analysis show that a fibre-matrix interface failure initiates by onset of a
finite debond in the neighbourhood of an
  interface point  where the failure criterion is reached first (under increasing proportional load), this debond further propagating along the interface in   mixed   mode  or even, in some configurations,  with the crack tip   under compression.
  The analytical  predictions of the  debond onset position    and  associated critical load are used for checking the computational procedure implemented, an excellent agreement being obtained.
 }

\end{abstract}


\end{frontmatter}


\section{Introduction}
\label{Introduction}
 Matrix  (or interfibre) failure in composite unidirectional laminates
subjected to loads transverse to the fibres is often initiated by the debonding of some fibres~\cite{HullClyne,Zhang1997,Varna1997,ParisCorrea2007,Correa2008a,Correa2008b}. The 
problem of an
elastic circular cylindrical inclusion (fibre) embedded in an elastic matrix without or with a partial debond
at their interface, subjected to uniaxial tensile/compressive loads,
has intensively been  studied in the past. An extensive review of these works can be found in~\cite{ParisCorrea2007,mantic2009,TavaraEABE2011}.
In the present work,  debond onset and propagation   along the interface of an isolated fibre embedded in
an elastic matrix subjected to remote biaxial transverse loads is studied,  cf.~\cite{ParisCorrea2003,manticgarcia2012,Correa2013}. The aim  is to obtain,
   among other  results,  failure curves predicting the critical loads which   cause  the fibre-matrix interface failure. The  results presented
may contribute to understand the mechanisms of damage initiation in unidirectional composite laminas under transverse loads.

In many practical situations, the behavior of (adhesively) bonded solids can be described by modeling
a thin (adhesive) elastic layer, also called interphase,  as a continuos distribution of linear-elastic
springs with appropriate stiffness parameters~\cite{Goland1944,erdogan1997,geymonat1999,benveniste2001,hashin2002,lenci2001}.
This classical model is usually referred to as \emph{linear-elastic interface, weak interface} or \emph{imperfect interface}.
As proposed recently by several authors~\cite{TavaraEABE2011,Caporale2006,Bennati2009,Cornetti2009,TavaraCMES2010,Cornetti2012,Becker2013}, a practical way to describe  debonding or delamination processes is to enrich this classical model by strength  and fracture parameters and associated failure criteria. Such a model is considered as a limit ``non-smooth case'' of some (nonlinear) cohesive zone models (CZMs) in~\cite{Valoroso2006}.

With reference to the particular problem of   fibres embedded in a matrix, many authors
consider that an appropriate manner to describe the physical nature and mechanical
behavior of the fibre-matrix interface is by applying this  elastic interface  model, see~\cite{TavaraEABE2011,hashin2002} and references therein. An analytical closed-form solution of a single circular inclusion problem, assuming an
undamaged linear-elastic interface under remote tension was deduced by Gao~\cite{Gao1995}.
A generalization of this solution  was later presented by Bigoni \textit{et al.}~\cite{bigoni1998}.
 Mogilevskaya and Crouch~\cite{Mogilevskaya2002} solved numerically the problem
of an infinite, isotropic elastic plane containing a large number of randomly distributed circular
elastic inclusions with spring-like interface conditions. Later, Caporale \textit{et al.}~\cite{Caporale2006}
applied a linear elastic - (perfectly) brittle law, using normal and shear interface-strength criteria, and the 3D finite element method (FEM)
to determine curves of macro-strains corresponding to the initiation of the interfacial debonding.

Following similar ideas, other authors have applied different CZMs to model the fibre-matrix debond, a few of them being mentioned herein.   Levy and co-workers in a series of works, see~\cite{XieLevy2007} and references therein, 
carried out   parametric studies of the stability of the phenomenon of circular-inclusion decohesion under biaxial loading applying a CZM. Carpinteri \textit{et al.}~\cite{Paggi2005}  used a CZM    to study the instability phenomena in fibrous metal matrix composites by FEM.
Han \textit{et al.}~\cite{Han2006} used    a softening decohesion model to study the initiation and propagation of debonds in several single and two fibre configurations by the boundary element method (BEM). Recently, Ngo \textit{et al.}~\cite{Paulino2010}
used a new potential-based CZM to study the inclusion-matrix debonding in an integrated approach involving micromechanics, and Kushch \textit{et al.}~\cite{Kushch2011} used a bi-linear CZM  to simulate progressive debonding in  multi-fiber models of a composite showing  formation of debond clusters.

An alternative analytical approach based on a coupled stress and energy criterion \cite{mantic2009,leguillon2002} and the classical open model of interface cracks \cite{germ2006} has recently been applied by  Manti\v{c} and Garc\'{\i}a~\cite{manticgarcia2012} to characterize  the   initiation and propagation of a fibre-matrix interface crack  under biaxial loads.

In the present work, the original linear elastic - (perfectly) brittle interface  model (LEBIM) developed by T\'{a}vara \textit{et al.}~\cite{TavaraEABE2011,TavaraCMES2010} is  employed because of its simplicity, robustness and computational efficiency. This model is enhanced by considering the possibility of   frictionless elastic contact at broken portions of the interface and also  by extending the range of  variation of the interface fracture toughness with the fracture mode mixity.
This new LEBIM is used together with Gao's  analytical solution \cite{TavaraEABE2011,Gao1995} for evaluating a failure curve of a single fibre  under biaxial loads, which may provide  an approximation of the corresponding failure curve for  dilute fibre packing (low fibre volume fraction).
The   LEBIM is also  implemented in a 2D collocational BEM code, 
used to study the debond initiation and propagation in the present work, and  will allow   solving accurately and efficiently   the problem  of debond initiation and propagation for dense fibre packing (high fibre volume fraction) including many fibres in forthcoming works, see~\cite{Tavaratesis,TavaraBETEQ2013} for some preliminary results.

The LEBIM with the extended interface failure criterion is presented in Section 2. In Section 3, the problem of a circular inclusion under a remote biaxial   transverse loading is defined and Gao's  analytical solution  is   reviewed.  Both the analytical and numerical BEM procedures  for the fibre-matrix debond modeling are described in  Section 4.  Finally, the influence of the three dimensionless governing parameters (ratio of the interface shear and normal stiffnesses $\xi$, fracture mode-sensitivity parameter $\lambda$ and brittleness number $\gamma$), in addition to the load biaxiality parameter $\chi$, on  the    position of debond initiation, value of   critical biaxial transverse load and    further debond propagation is studied and discussed in Section 5.

\section{Linear elastic - (perfectly) brittle interface model (LEBIM)}
\label{LEBIsec}

New enhanced constitutive law and failure criterion of the LEBIM, cf. \cite{TavaraEABE2011,TavaraCMES2010,Tavaratesis}, are introduced in this section. Although this interface model is originally considered   representing an adhesive layer of a small thickness $h>0$,   it can be applied to simulate  debonding mechanisms of   bimaterial systems where,  strictly speaking, there is no additional third material between bonded materials, as may occur in the present case  of  fibre-matrix interface in a real composite. Actually, the continuous distribution of springs in the LEBIM has zero thickness.

\subsection{Constitutive law of the spring distribution}

\begin{figure}[!htb]
\centering
\mbox{Linear elastic - (perfectly) brittle interface}\\[0.1em]
\subfigure[$\delta_n(x) \leq \delta_{nc}(\psi)$]{\includegraphics[angle=270,scale=0.25]{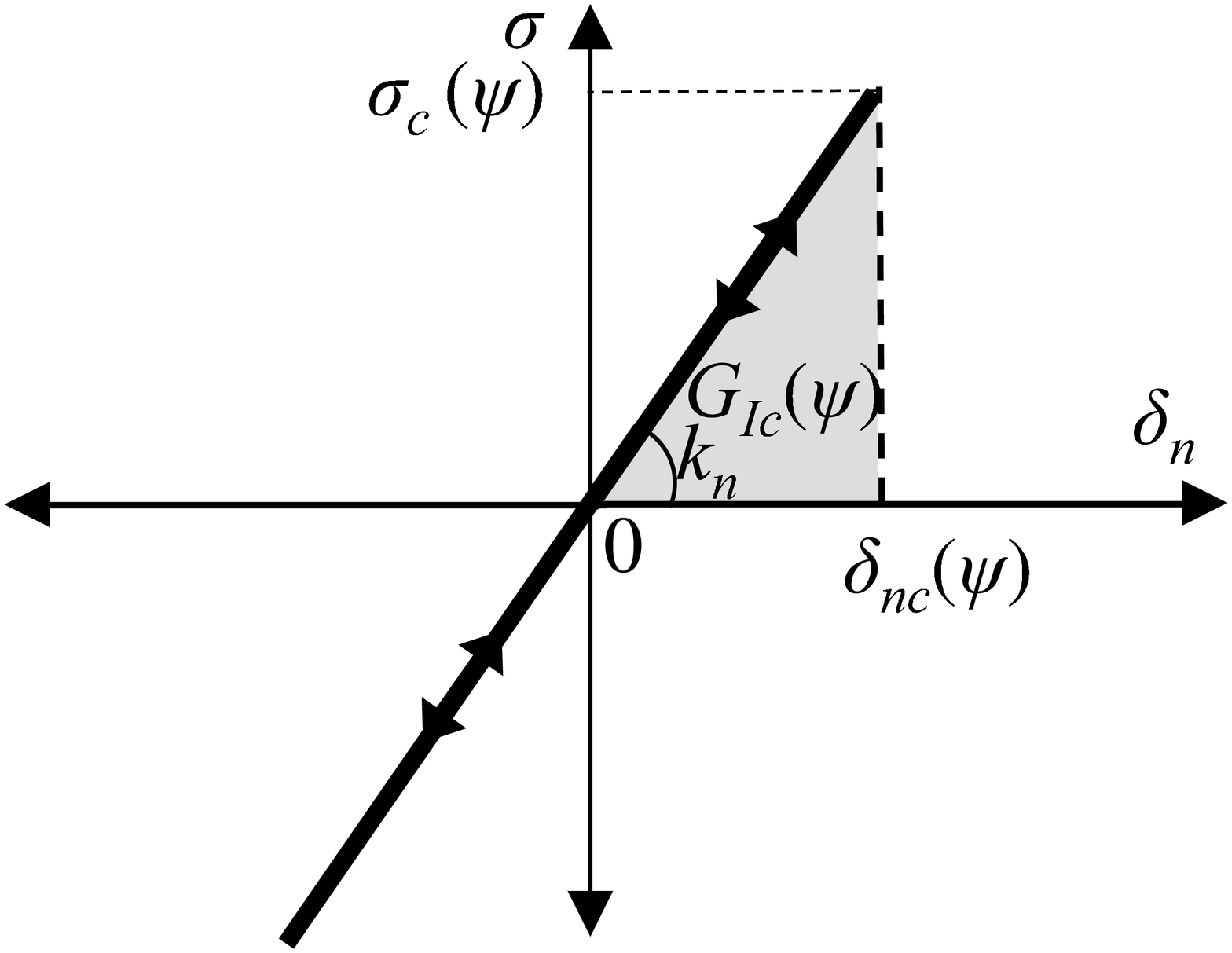}}\hspace{0.1cm}
\subfigure[$|\delta_t(x)| \leq \delta_{tc}(\psi)$]{\includegraphics[angle=270,scale=0.25]{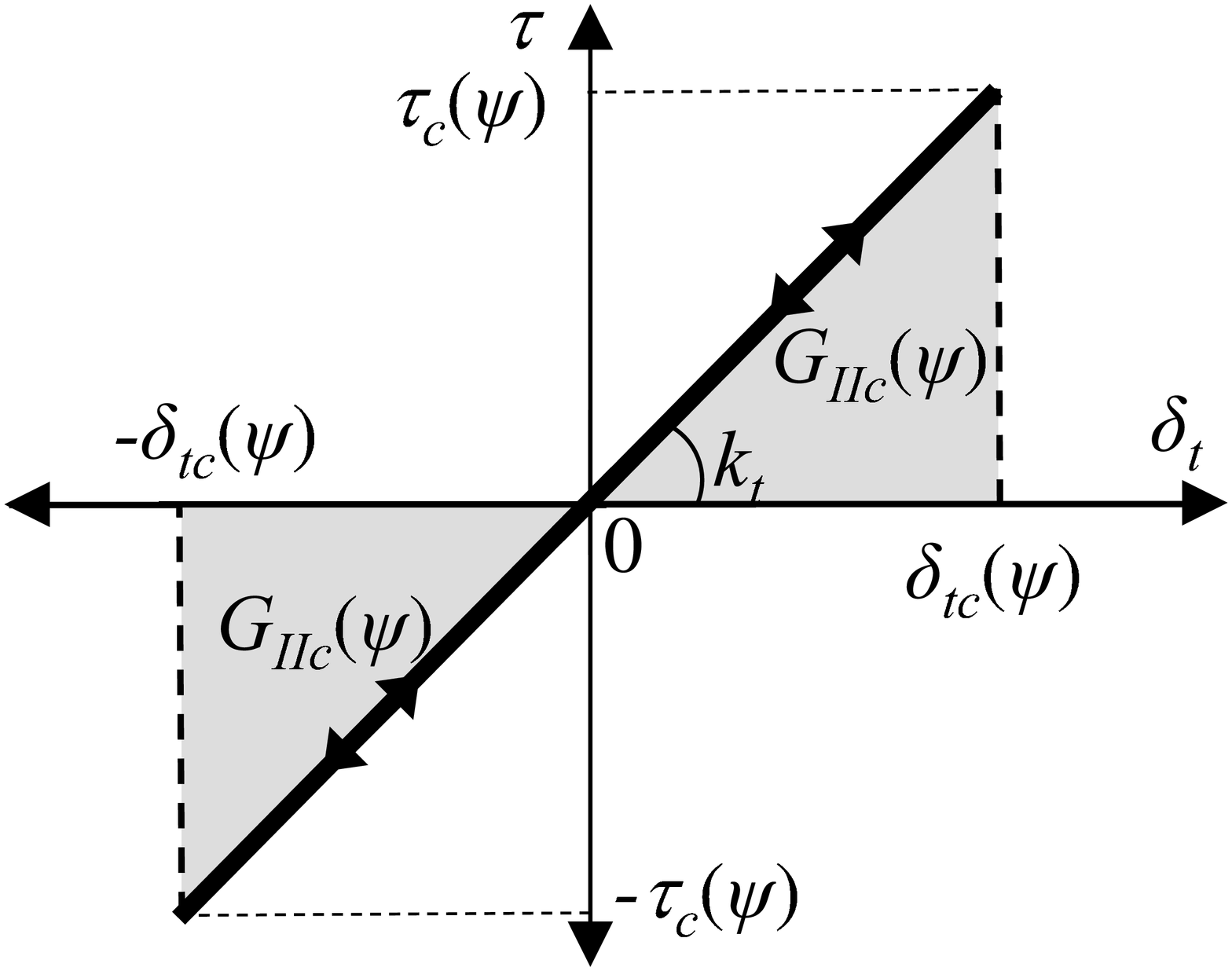}}\hspace{0.1cm}
\\[1em] \mbox{Broken interface}\\[0.1em]
\subfigure[]{\includegraphics[angle=270,scale=0.25]{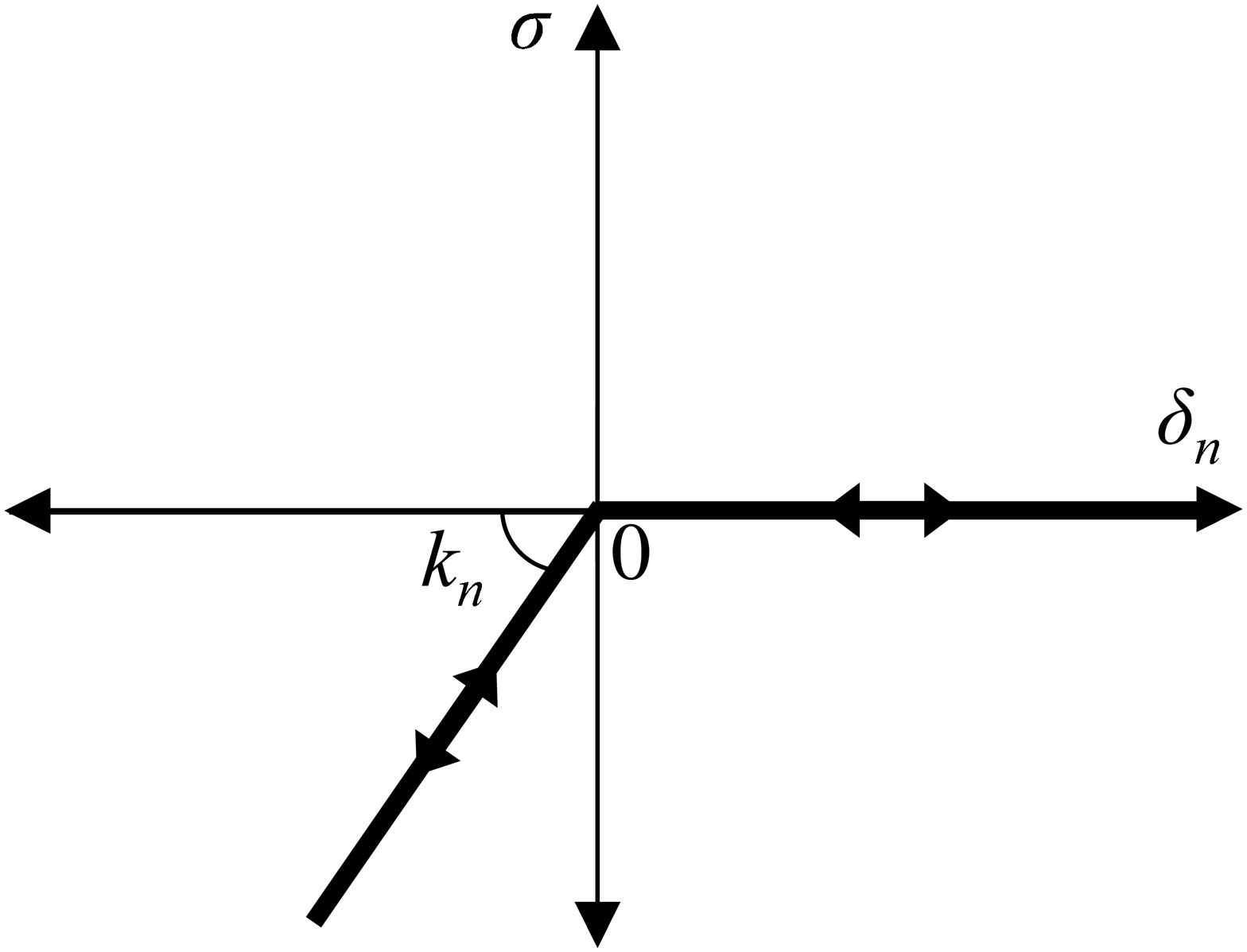}}\hspace{0.1cm}
\subfigure[]{\includegraphics[angle=270,scale=0.25]{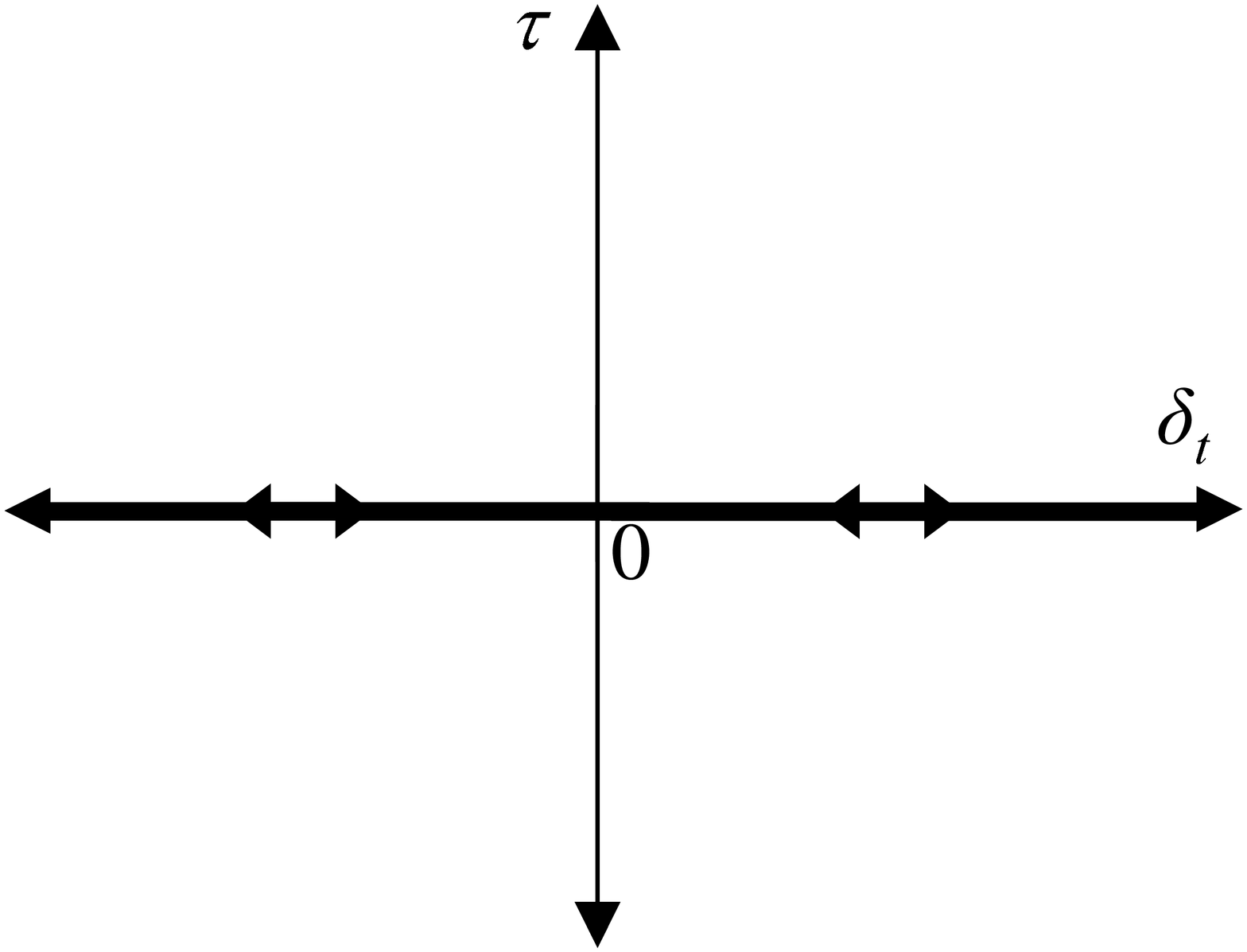}}
\caption{ Linear elastic - (perfectly) brittle law in the undamaged interface  in the (a) normal and (b) tangential directions, and in  the broken interface  in the (c) normal and (d) tangential directions.}\label{laws}
\end{figure}

The constitutive law of the continuous spring distribution is defined by a relation between tractions and relative displacements at the interface, Fig.~\ref{laws}.
When modeling an undamaged isotropic layer this spring distribution is governed by the following simple linear-elastic
law  written at an interface point $x$, Fig.~\ref{laws}(a) and (b):
\begin{equation}
\begin{array}{lll}
\begin{array}{l}
  \textrm{Linear Elastic} \\
  \textrm{Interface}
\end{array}  &
         \left\{\begin{array}{l}
            \sigma(x)=k_n\delta_n(x), \\
            \tau(x)=k_t\delta_t(x),
          \end{array}\right.  \quad \mathrm{for} \quad
          & t(x)<t_c(\psi(x))
\label{eqlaw1}
\end{array}
\end{equation}
where $\sigma(x)$ and $\tau(x)$ are the normal and tangential tractions
at a point $x$ of the elastic layer, $\delta_n(x)$ and $\delta_t(x)$  are the normal (opening) and tangential (sliding)
relative displacements between opposite interface points, and
$k_n$ and $k_t$ denote the normal and tangential stiffnesses of the spring distribution,  respectively.

The failure criterion may be written in terms of the traction modulus at every point $x$,
$t(x)=\sqrt{\sigma^2(x)+\tau^2(x)}$. The interface  breaks at a point $x$ when the traction modulus $t(x)$ reaches
its critical value:
\begin{equation}
 t_c(\psi(x)) = \sqrt{\sigma^2_{c}(\psi(x))+\tau^2_{c}(\psi(x))},
\label{tracm}
\end{equation}
where the critical normal and tangential tractions $\sigma_{c}(\psi(x))$ and
$\tau_{c}(\psi(x))$, and the corresponding critical relative displacements $\delta_{nc}(\psi(x))$ and
$\delta_{tc}(\psi(x))$, are functions of the fracture-mode-mixity angle $\psi$
at a particular point $x$. Thus, different critical values of these variables may be
obtained at different interface points, due to the fact that $\psi$ can vary along
the adhesive layer.

Once the failure criterion in \eqref{eqlaw1}, which will be described in the following section, is reached the damaged interface
is considered free of stresses, unless contact appears between both sides of the damaged interface.
In this case, the interface retains its normal stiffness. Therefore, once the interface is broken,
the following non-linear constitutive law\footnote{Let us recall the definition of the  positive and negative part of a real number $\delta$ used in the present work, $\langle \delta\rangle_{\pm}=\frac{1}{2}\left(\delta\pm|\delta|\right)$. 
$\langle \cdot \rangle_{+}$  is also referred to as Macaulay brackets or ramp function.}
is considered at an interface point $x$, Fig.~\ref{laws}(c) and (d):
%
\begin{equation}
\begin{array}{lll}
       \begin{array}{l}
  \textrm{Broken} \\
  \textrm{Interface}
\end{array}  & \left\{\begin{array}{l}
                                       \sigma(x)= k_n\langle \delta_n(x)\rangle_{-},  \\
                                       \tau(x)=0.
                                     \end{array}\right.
       \end{array}
\label{eqlaw2}
\end{equation}

\begin{figure}[!htb]
\centering
\subfigure[]{\includegraphics[clip=true,angle=270,scale=0.26]{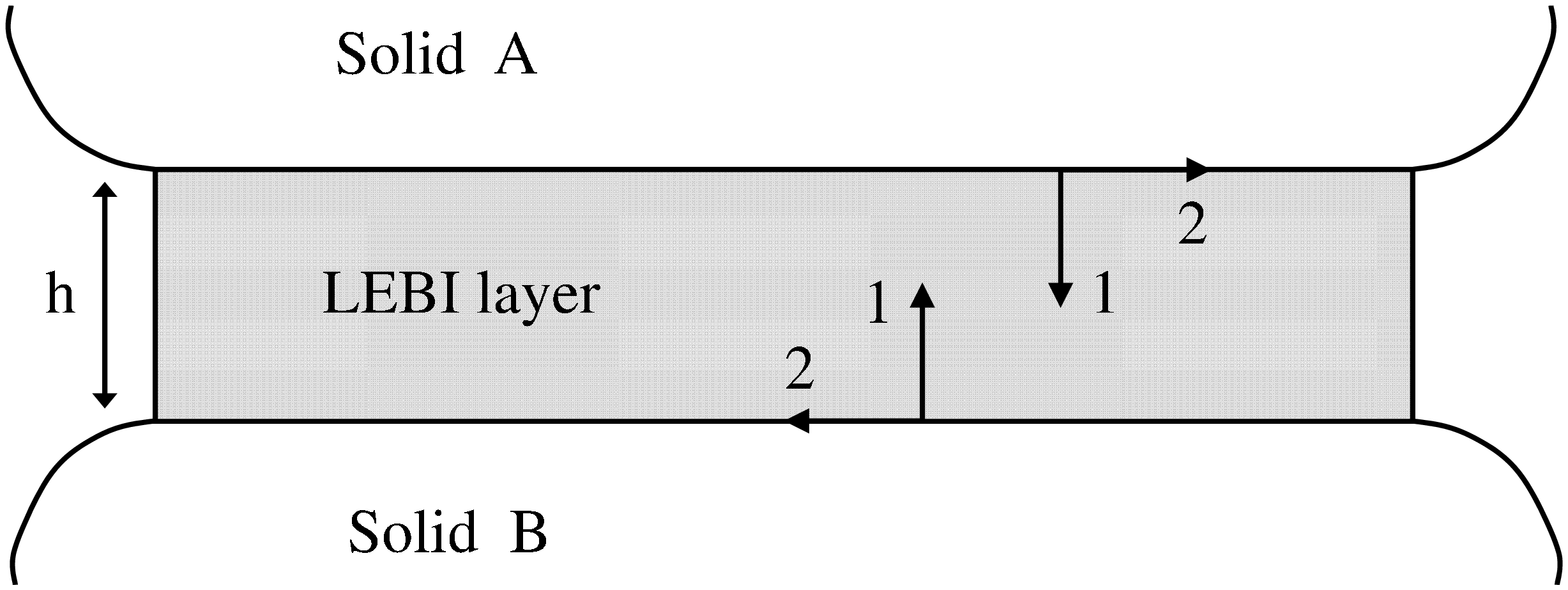}}\hspace{0.2cm}
\subfigure[]{\includegraphics[clip=true,angle=270,scale=0.26]{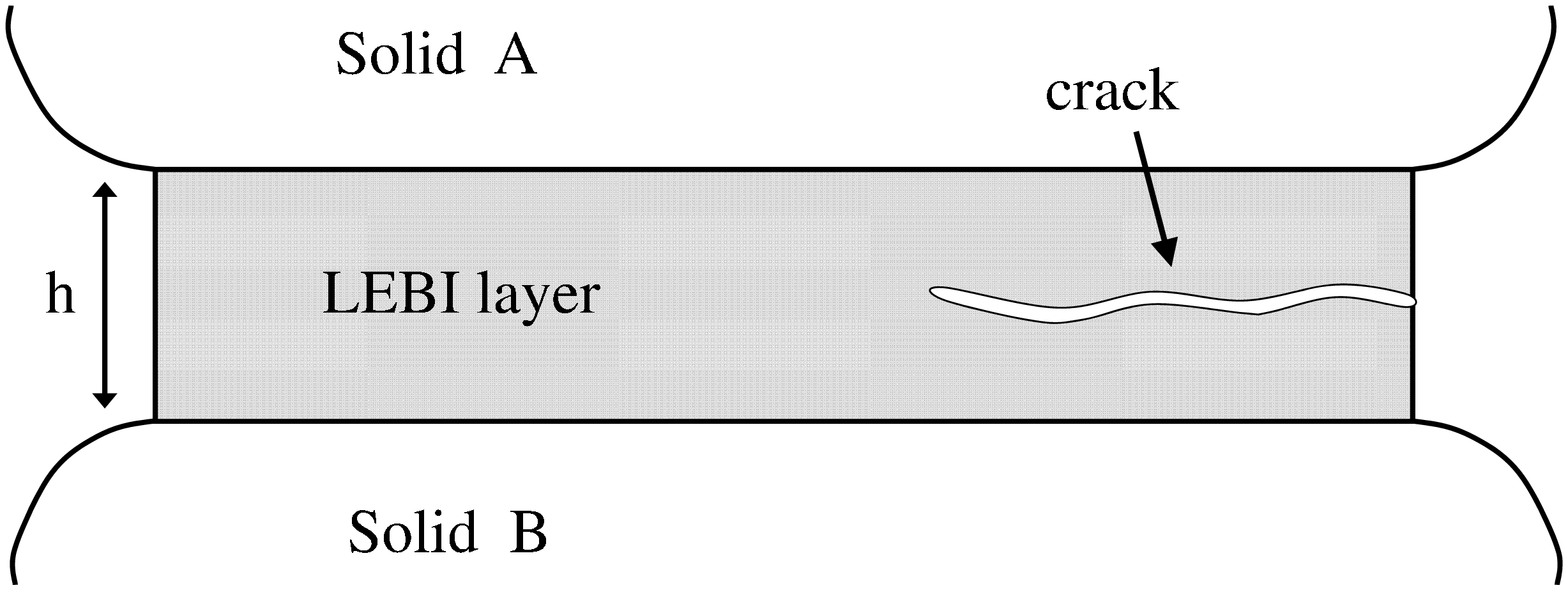}}
\caption{LEBI layer (a) undamaged and (b) partially broken.}
\label{lebicontact}
\end{figure}

Regarding the normal linear elastic - (perfectly) brittle law, once a portion of
interface is cracked,  large negative values of the normal relative displacement, $\delta_n<0$, are essentially
avoided by using the frictionless contact
condition \eqref{eqlaw2}, see Fig.~\ref{laws}(c).
The use of an elastic frictionless contact is based on the idea that   some portions of the cracked layer remain  on the
adjacent surfaces. Thus, when these surfaces enter in contact, it seems reasonable to assume that
these portions of the layer could compress with the same stiffness in the normal direction as the layer had before
cracking, see Fig.~\ref{lebicontact}.

The stiffness parameters
$k_n$ and $k_t$ could be related to the parameters of a   linear elastic isotropic  layer (Young's modulus $E_{\ell}$,
Poisson's ratio $\nu_{\ell}$, shear modulus $\mu_{\ell}$,
Lame's parameter $\lambda_{\ell}$,
and a small thickness $h$)~\cite{TavaraEABE2011} by:
\begin{equation}
k_n=\frac{2\mu_{\ell}+\lambda_{\ell}}{h}=\frac{E'_{\ell}}{h(1-\nu_{\ell}'^2)}=\frac{E_{\ell}(1-\nu_{\ell})}{h(1+\nu_{\ell})(1-2\nu_{\ell})},
\label{kn}
\end{equation}
\begin{equation}
k_t=\frac{\mu_{\ell}}{h}.
\label{eqcubo2}
\end{equation}
where $E_{\ell}'=E_{\ell}/(1-\nu_{\ell}^2)$ and $\nu_{\ell}'=\nu_{\ell}/(1-\nu_{\ell})$.
From \eqref{kn} and \eqref{eqcubo2} the following expression of the ratio of $k_t$ and $k_n$  can be obtained:
\begin{equation}
\xi=\frac{k_t}{k_n}=\frac{\mu_{\ell}}{2\mu_{\ell}+\lambda_{\ell}}=\frac{1-\nu_{\ell}'}{2}=\frac{1-2\nu_{\ell}}{2(1-\nu_{\ell})}
\label{knkt}
\end{equation}
leading to the following constraint for thin isotropic layers $0\leq \xi=k_t/k_n \leq 0.5$.

\subsection{Interface failure criterion}

The interface failure criterion is based on the Energy Release
Rate (ERR) concept, although its final expression used in the computational
implementation is given in terms of the interface tractions. As  the LEBIM implies the absence of stress singularities
at the crack tip, the ERR in a linear interface model is defined as the stored
elastic strain energy per unit length in the
unbroken ``interface spring'' at the crack tip (infinitesimal interface segment situated
at the crack tip) \cite{lenci2001,Cornetti2009}. Thus, the ERR of a mixed mode crack
in a linear elastic   interface is defined as, cf.~\cite{TavaraEABE2011,TavaraCMES2010}:
\begin{equation}
G=G_I+G_{II}\quad \textrm{ with } \quad G_I=\frac{\sigma \langle\delta_n\rangle_{+}}{2} \quad \textrm{and} \quad  G_{II}=\frac{\tau \delta_t}{2},
\label{Gdef}
\end{equation}
verifying $G_{I}=0$ for $\delta_n\leq 0$.

An extension of the energetic fracture-mode-mixity-angle $\psi_G$, introduced in~\cite{TavaraEABE2011,TavaraCMES2010} by the relation $\tan^2\psi_G=G_{II}/G_{I}$ for $G_{I}> 0$,   which will cover  also an interface under compression with $\sigma<0$, can be defined by
\begin{equation}\label{def_psi}
\tan \psi=\sqrt{\xi^{-1}}\tan\psi_\sigma=\sqrt{\xi}\tan\psi_u,
\end{equation}
where $\tan\psi_\sigma=\tau/\sigma$ and $\tan\psi_u=\delta_t/\delta_n$, $\psi_\sigma$ and $\psi_u$ being the stress and relative displacement based fracture-mode-mixity angles, respectively. Notice that $\psi=\psi_G$ for $\sigma>0$, and that absolute value of tangent of $\psi$  is given by the geometric mean of tangents of $\psi_\sigma$ and $\psi_u$, i.e. $|\tan\psi|=\sqrt{\tan\psi_\sigma\tan\psi_u}$.

According to the interface failure criterion proposed in the present work, an  
interface point breaks  when the  ERR $G$ reaches the fracture energy $G_c(\psi)=G_{Ic}(\psi)+G_{IIc}(\psi)$ (cf. Fig.~\ref{laws}(a) and (b)), which depends on the fracture mode mixity, i.e. $G = G_c(\psi)$.  By a suitable modification of the phenomenological law $G_c(\psi)=\bar{G}_{Ic}[1+\tan^2((1-\lambda)\psi)]$, suggested in \cite{HS92}, the following general expressions of the critical values of interface normal and tangential tractions as well as of the   normal and tangential relative displacements (shown in Fig.~\ref{laws})
as  functions of the fracture-mode-mixity-angle $\psi$ are obtained, cf. \cite{TavaraEABE2011,Tavaratesis}:
\begin{subequations}
\label{interfacelaw}
\begin{align}
\sigma_c(\psi) &=\bar{\sigma}_c\hat{\sigma}_c(\psi) =\bar{\sigma}_c\sqrt{1+\tan^2 [(1-\lambda)\psi]}\, \cos\psi, \quad
  &\delta_{nc}(\psi)=\frac{\sigma_c(\psi)}{k_n},
\label{sigmac}\\
\tau_{c}(\psi) &=\bar{\sigma}_c\hat{\tau}_c(\psi)   =\bar{\sigma}_c\sqrt{\xi}\sqrt{1+\tan^2 [(1-\lambda)\psi]}\, \sin\psi,\quad
 &\delta_{tc}(\psi)=\frac{\tau_c(\psi)}{k_t},
\label{tauc}
\end{align}
\end{subequations}
where $\bar{G}_{Ic}$ is the interface fracture toughness in pure mode I, $\bar{\sigma}_c>0$ is the critical interface normal  stress in pure mode I (interface tensile strength) and $\lambda$ $(0\leq \lambda\leq 1)$ is a fracture mode-sensitivity parameter obtained experimentally.  A typical range
$0.2 \leq \lambda \leq 0.3$ characterizes interfaces with moderately strong fracture mode dependence \cite{HS92}.

It should be noticed that if $\bar{G}_{Ic}$ and $\bar{\sigma}_c$ values are obtained experimentally,
then $k_n$ is given by the relation $\bar{G}_{Ic}=\bar{\sigma}_c^2/2k_n$.
Thus, 
the LEBIM needs the input of four
independent variables: $\bar{G}_{Ic}$, $\bar{\sigma}_c$, $\xi$ and $\lambda$.

\begin{figure}[!htb]
\centering
\includegraphics[angle=270,scale=0.32]{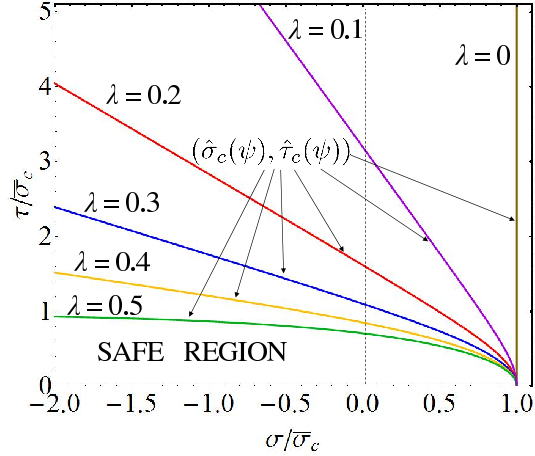}
\caption{Interface failure curve in   plane $(\sigma/\bar{\sigma}_c,\tau/\bar{\sigma}_c)$ for different values of $\lambda$ and $\xi=k_t/k_n=0.25$.}
\label{sigmataulam}
\end{figure}

The plot of the interface failure curve parameterized by equations \eqref{interfacelaw}, in the plane of normalized interface stresses $(\sigma/\bar{\sigma}_c,\tau/\bar{\sigma}_c)$,  considering $\xi=k_t/k_n=0.25$, is shown in Fig.~\ref{sigmataulam} where only the upper half of these curves is plotted  for $\tau\geq 0$.
According to Fig.~\ref{sigmataulam}, an interface failure  under  compressions is possible but requires  larger shear stresses. As a consequence, a closed crack with compressions in the neighbourhood of the crack tip   may propagate in presence of sufficiently large shear stresses.

The interface failure curves  for $0\leq\lambda\leq 0.5$  are open having two asymptotes whose angles are
\begin{equation}\label{psia}
\psi_a=\frac{\pi}{2(1-\lambda)}
 \end{equation}
 and $-\psi^a$, see Fig.~\ref{sigmataulam},   an interface failure for these values of $\lambda$ being  only possible for  $|\psi|<\psi_a$. It is easy to see that  $G_c(\psi)$ is unbounded  for  $\psi$ approaching $\psi_a$ $(0\leq\lambda<0.5)$ \cite{TavaraEABE2011,Tavaratesis,HS92}. Notice that these interface failure curves are closed for $\lambda>0.5$, reducing to an ellipse for $\lambda=1$.

Failure (damage) of a portion of the interface layer is
modeled as an abrupt decrease (jump down) of stresses in this zone of the layer, associated to a
free separation or sliding of both interface surfaces, when a point on the failure curve
(in $(\sigma/\bar{\sigma}_c,\tau/\bar{\sigma}_c)$ plane) is achieved in that portion of the layer. Actually, in view of Fig.~\ref{laws}, in the interface portion under compression only shear stresses jump down after its failure.

It is noteworthy that Fig.~\ref{sigmataulam} reminds other   interface failure criteria as those presented
by Lemaitre and Desmorat in \cite{LemaitreDesmorat2005} (Fig.~7.5 therein) and by Bialas and Mr\'oz in \cite{BialasMroz2005} (Fig.~3 therein), although based on different approaches.

\section{Problem of a circular inclusion under biaxial transverse loads}\label{Section_Problem}

\begin{figure}[!htb]
\centering
\subfigure[]{\includegraphics[angle=270,scale=0.35]{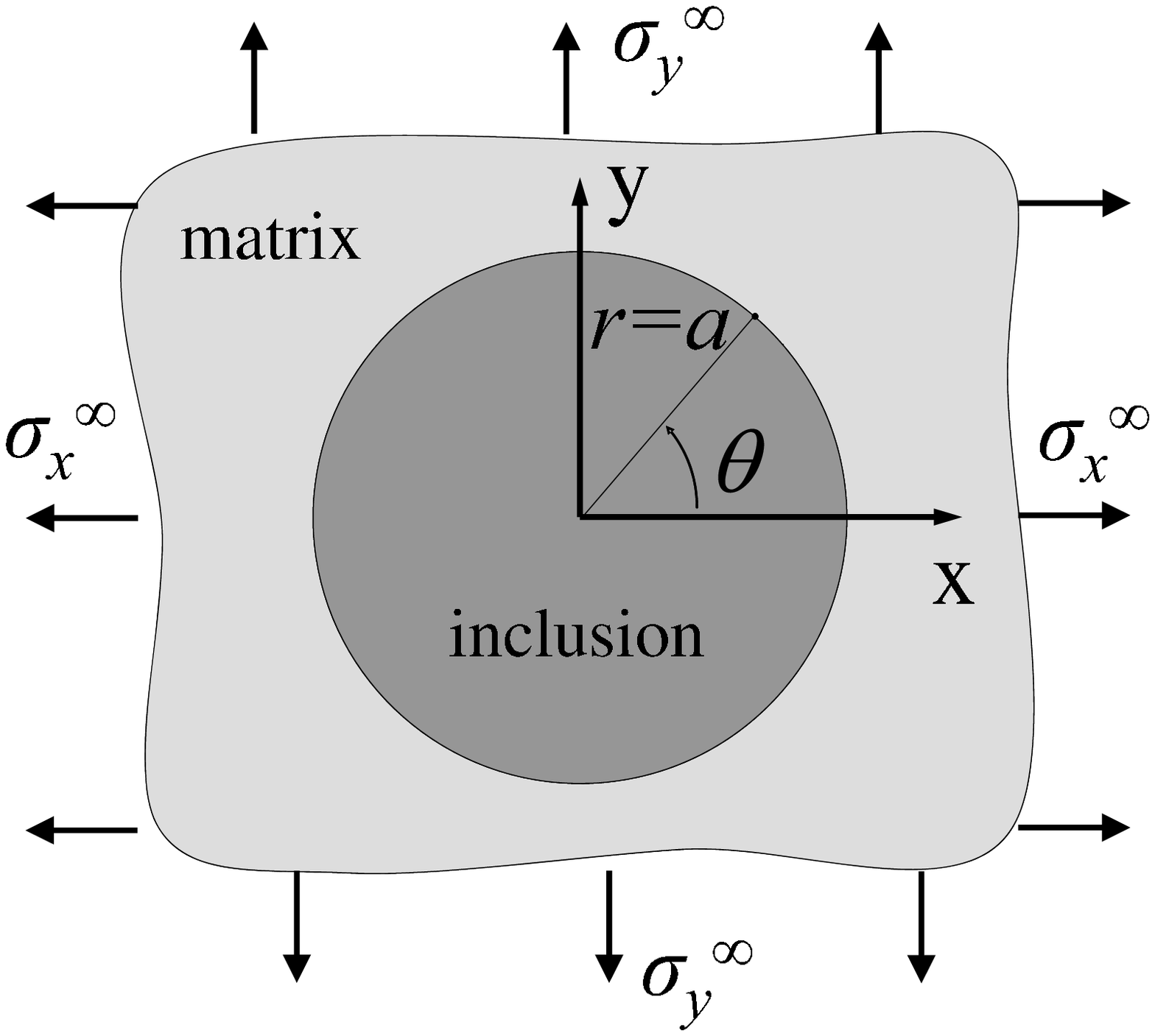}}\hspace{1.5cm}
\subfigure[]{\includegraphics[angle=270,scale=0.35]{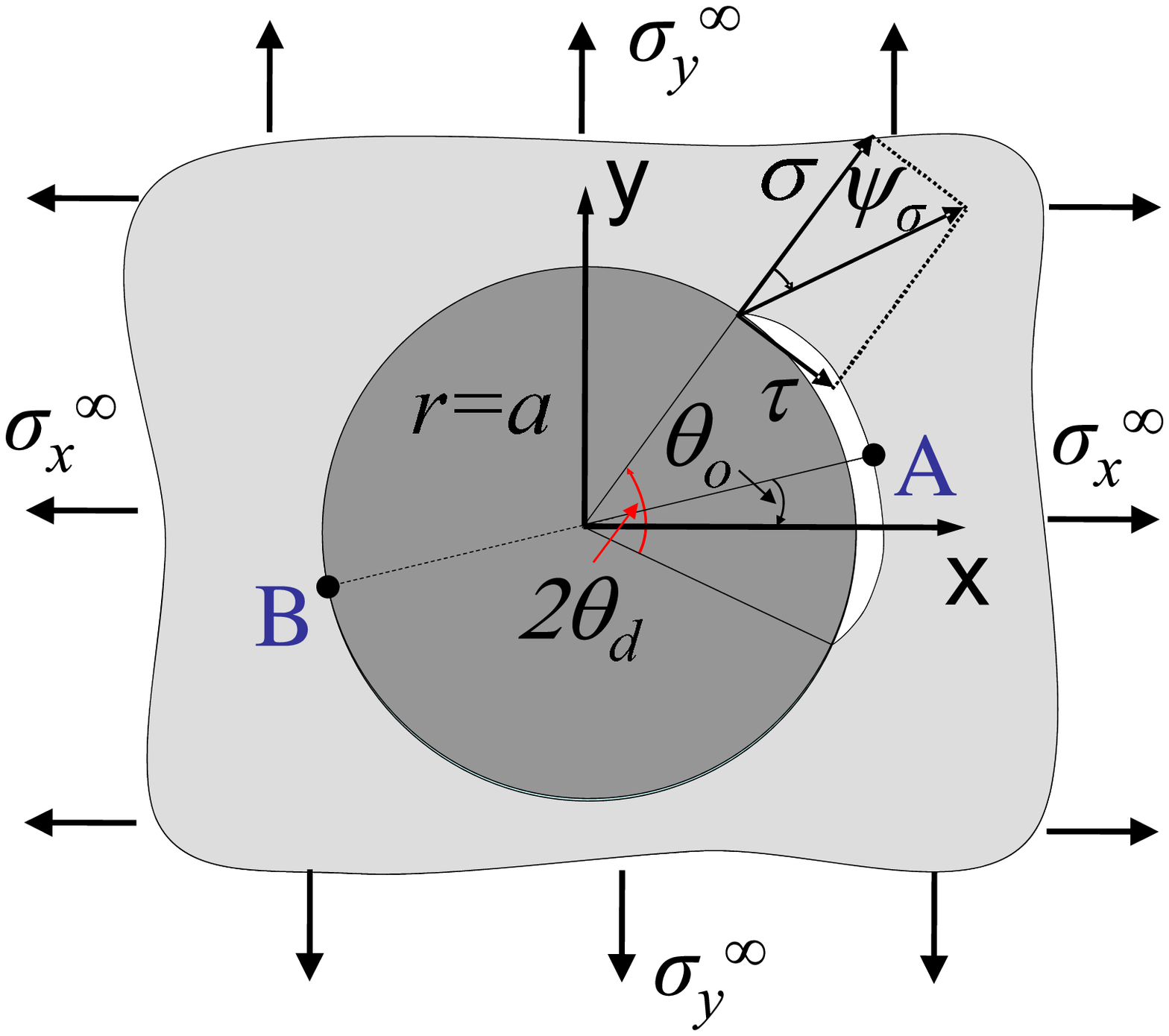}}
\caption{Inclusion problem configuration under biaxial remote transverse tension (a) without and (b) with a partial debond.}
\label{fibmat}
\end{figure}

The plane strain problem of a circular inclusion of radius $a>0$  embedded in an  infinite   matrix,
initially without any debond along its interface, and subjected to   remote
uniform   stresses is considered. The materials of both the inclusion and   matrix are considered to be linear elastic isotropic.
Let $(x, y)$ and $(r,\theta)$ be the cartesian and polar coordinates with the origin of coordinates in the center of the inclusion, assuming without any loss of generality that $(x, y)$ is the principal coordinate system of the remote stress state defined by the  principal stresses
$\sigma_x^{\infty} \ge  \sigma_y^{\infty}$, see Fig.~\ref{fibmat}.

 Although  the  ratio of the  principal stresses $\eta=\sigma_y^{\infty}/\sigma_x^{\infty}=\tan\phi^\infty$ is sometimes used to characterize the biaxiality of the  remote stress state \cite{manticgarcia2012}, in the present work, which covers also configurations where both remote principal stresses are compressive, the following general \emph{load-biaxiality parameter}\footnote{It is easily to see that $\chi$ gives the position of the center of the normalized Mohr circumference and its characteristic values are $\chi=1$ -  equibiaxial  tension, $\chi=0.5$ - uniaxial tension, $\chi=0$ - equibiaxial  tension-compression (pure shear stress), $\chi=-0.5$ - uniaxial compression and  $\chi=-1$ -  equibiaxial  compression). It is useful to realize that $\phi^\infty=\frac{\pi}{2}\left(\chi-\frac12\right)$.}:
\begin{equation}\label{chidef}
\chi=\frac{\sigma_x^{\infty}+\sigma_y^{\infty}}{2\max\{|\sigma_x^{\infty}|,|\sigma_y^{\infty}|\}},\qquad -1\leq\chi\leq 1
\end{equation} is more suitable. Denoting the Frobenius norm of the remote stress state   by $S^\infty=\sqrt{(\sigma_x^{\infty})^2+(\sigma_y^{\infty})^2}$, we have
 $\sigma_x^{\infty}=S^\infty\cos\phi^\infty$ and
$\sigma_y^{\infty}=S^\infty\sin\phi^\infty$.

Let the position where the interface crack onset occurs be defined by the polar angle $\theta_o\in\langle 0^\circ,90^\circ\rangle$. The semidebond angle is denoted as $\theta_d$. During the debond growth the angle  $\theta_o$ may or may not be placed at the center of the debond.

According to Fig.~\ref{fibmat}(b) only one debond, initiated at a point A($r=a$, $\theta=\theta_o$), is considered, although  depending on the problem symmetry two or four equivalent positions for debond onset may exist at the inclusion interface with $\theta=\pm\theta_o$, $\pm\theta_o+180^\circ$.
   Nevertheless, according to the experimental evidence only one side of the
fibre-matrix interface is usually broken \cite{Zhang1997,CorreaGamstedt2007}. This will also be obtained by the present numerical model in Section~\ref{SectionResults},  where the crack onset can occur at any of these two or four points, but
once a crack has started at one of these points it will continue growing, preventing failure in the other symmetrically situated points.

A typical bi-material system among fibre reinforced composite materials is chosen for this
study: $m$-epoxy matrix and $i$-glass fibre (inclusion), the elastic properties of matrix and fibre being
$E_m=2.79 $ GPa, $\nu_m=0.33$, $E_i=70.8 $ GPa and $\nu_i=0.22$, respectively.
The corresponding Dundurs bi-material parameters in plane strain are $\alpha=0.919$ and $\beta=0.229$  and the harmonic mean of the effective elasticity moduli is $E^*=6.01$GPa,  see \cite{ParisCorrea2007,mantic2009,germ2006,Soden,Schulte} for their definitions.

The  strength and fracture properties   of the fibre-matrix interface, tensile strength $\bar{\sigma}_c=90$ MPa and fracture energy in mode I  $\bar{G}_{Ic}=2$ Jm$^{-2}$, considered in the numerical procedure are in the range of values found in the literature \cite{Zhang1997,Varna1997}, and correspond to quite  brittle behaviour \cite{mantic2009,manticgarcia2012} making the hypothesis of the LEBIM to represent appropriately a possible real composite material behavior \cite{TavaraEABE2011}.

A dimensionless structural parameter, referred to as brittleness number,  governing brittle-to-tough transition in the fibre-matrix debond onset can be defined following \cite{mantic2009,TavaraEABE2011,manticgarcia2012} as
\begin{equation}\label{gammadef}
\gamma=\frac{1}{\bar{\sigma}_c}\sqrt{\frac{\bar{G}_{Ic}E^*}{a}}=\sqrt{\frac{E^*}{k_n2a}},
\end{equation}
where the second expression, showing that $\gamma$ is given by the ratio of stiffnesses of the bimaterial ($E^*$) and interface ($k_n$) with the unique characteristic length of problem
geometry (fibre diameter $2a$), is obtained using the relation $\bar{G}_{Ic}=\sigma_c(0^\circ)\delta_{nc}(0^\circ)/2=\sigma^2_c(0^\circ)/2k_n$ \cite{TavaraEABE2011,TavaraCMES2010}. Small values of $\gamma$ (typically $\gamma\lesssim 1$) correspond to brittle and large values of $\gamma$  (typically $\gamma\gtrsim 1$) to tough configurations. Noteworthy $\gamma$ is closely related to a similar dimensionless parameter $\delta$ defined by Lenci \cite{lenci2001} for a crack of size $2a$ at a weak interface, verifying  $\gamma\sim 1/\sqrt{\delta}$.

In the following numerical study, some parametric analyses will be presented, all of them consider a default configuration with $\xi$=0.25, $\lambda$=0.25 and a circular inclusion radius $a$=7.5 $\mu$m, leading to $\gamma=0.44$.

\section{Analytical and numerical procedures applied}
First, the analytical solution of the above defined problem of a circular inclusion (fibre) under remote biaxial transverse loads,    considering the inclusion-matrix interface  as a linear-elastic layer without any debond, is presented and discussed. Then based on this solution and the hypotheses of the LEBIM, an analytical procedure able to evaluate a  failure curve and the angle where debond onset takes place is proposed.  Finally a BEM model of this problem, able to analyse interface debond onset and propagation, is briefly described.

\subsection{Analytical procedure applied to analyse the fibre-matrix debond onset}\label{SectionAnalyticProcedure}
By using a closed-form expression of the Airy stress function deduced by Gao \cite{Gao1995} for an elastic circular inclusion (fibre) embedded in an elastic infinite matrix with an undamaged interface, the following expressions of interface tractions   can be obtained assuming     uniform biaxial stresses, $\sigma_x^{\infty}$ and $\sigma_y^{\infty}$, at infinity:
\begin{subequations}\label{Gaosolution}
\begin{align}
\sigma(r=a,\theta)&=\frac{k_n a (1+\kappa_m) }{2 A C}\{\sigma_x^{\infty}(A + B C \cos(2\theta))+\sigma_y^{\infty}(A + B C \cos(2(\theta+90^\circ)))\},
\label{sigmangao}\\
\tau(r=a,\theta)&=-\frac{k_t a (1+\kappa_m) D}{2 A}\{ \sigma_x^{\infty} \sin(2\theta)+\sigma_y^{\infty} \sin(2(\theta+90^\circ))\},
\label{sigmatgao}
\end{align}
\end{subequations}
where
\begin{subequations}\label{Gaoconstants}
\begin{equation}
A=12\mu_m^2+a^2k_n k_t(\kappa_m+t)(1+\kappa_i t)+a \mu_m(k_n+k_t)(1+3\kappa_m+(3+\kappa_i)t),
\label{Agao}
\end{equation}
\begin{align}
B&=6\mu_m+a k_t(1+\kappa_i t),
\label{Bgao}
\\
C&=4\mu_m+a k_n(2+(\kappa_i-1)t),
\label{Cgao}
\\
D&=6\mu_m+a k_n(1+\kappa_i t),
\label{Dgao}
\end{align}
\end{subequations}
with $t=\mu_m/\mu_i$, and   $\mu_m=E_m/2(1+\nu_m)$ and $\kappa_m=3-4\nu_m$, respectively
 being, the shear modulus   and Kolosoff constant   of the  matrix ($m$), and analogously for the inclusion ($i$).
 Equations \eqref{Gaosolution} and \eqref{Gaoconstants}   generalize
  expressions (27)-(31) introduced in \cite{TavaraEABE2011}\footnote{There are several misprints
in Eqs. (27)-(31) in \cite{TavaraEABE2011}:  in Eq. (28) the  minus sign is missing  and the term $B$
should be replaced by the missing term $D$ presented in  (\ref{Dgao}) herein, and the correct form of the term $B$ in Eq. (30) is given in \eqref{Bgao} herein.} for the uniaxial loading case ($\sigma_y^{\infty}=0$).

Taking into account that the parameters $A$, $B$, $C$ and $D$ can be written in terms of $\gamma$, $\xi$ and the elastic properties of matrix and inclusion \cite{TavaraEABE2011}, and that $k_na=E^*/2\gamma^2$ due to \eqref{gammadef}, the interface tractions in \eqref{Gaosolution} can be expressed in terms of dimensionless functions $\hat{\sigma}$ and $\hat{\tau}$  as:
\begin{subequations}\label{Gaonormalized}
\begin{align}
\sigma(r=a,\theta)&=S^\infty\, \hat{\sigma}(\theta;\chi,\xi,\gamma;E_i/E_m,\nu_i,\nu_m),\\
\tau(r=a,\theta)&=S^\infty \, \hat{\tau}(\theta;\chi,\xi,\gamma;E_i/E_m,\nu_i,\nu_m),
\end{align}
\end{subequations}
where $\xi$ \eqref{knkt}, $\chi$ \eqref{chidef} and $\gamma$ \eqref{gammadef} are the governing dimensionless parameters.

Pseudocode of the proposed   procedure for the evaluation of a failure curve in the plane of normalized remote stresses $(\sigma_x^\infty/\bar{\sigma}_c,\sigma_y^\infty/\bar{\sigma}_c)$, which uses the above analytical solution for interface tractions and assumes the hypotheses of the LEBIM, is introduced in Fig.~\ref{procedure}. Additionally, this procedure evaluates the polar angle $\theta_o$ where the debond initiates.  The procedure is self-explaining, thus  its detailed description is omitted for the sake of brevity.
\begin{figure}[!htb]
\noindent\small
\hspace*{\espacio em}\verb"Define "$\xi,\gamma,\lambda,E_i/E_m,\nu_i,\nu_m$\\
\hspace*{\espacio em}\verb"For "$\chi\in \langle -1,1\rangle$ \verb" Do"\\
\hspace*{\espacio em}\verb"   For "$\theta\in\langle 0,\frac{\pi}{2}\rangle$\verb" Do"\\
\hspace*{\espacio em}\verb"     Evaluate "$\hat{\sigma}$\verb" and "$\hat{\tau}(\theta;\chi,\xi,\gamma;E_i/E_m,\nu_i,\nu_m)$
\verb"  [Eqs."\eqref{Gaosolution}-\eqref{Gaonormalized}\verb"]"\\
\hspace*{\espacio em}\verb"     "$\hat{t}(\theta,\chi)=\sqrt{\hat{\sigma}^2(\theta,\chi)+\hat{\tau}^2(\theta,\chi)}$\\
\hspace*{\espacio em}\verb"     "$\psi(\theta,\chi)=\arctan\left(\sqrt{\xi}\hat{\sigma}(\theta,\chi),\hat{\tau}(\theta,\chi)\right)$
\verb"  [Eq."\eqref{def_psi}\verb"]" \\
\hspace*{\espacio em}\verb"     Evaluate "$\psi_a(\lambda)$\verb"   [Eq."\eqref{psia}\verb"]"\\
\hspace*{\espacio em}\verb"     If ("$0\leq\lambda\leq0.5$\verb" and "$\psi(\theta,\chi) < \psi_a(\lambda)$\verb") or "$0.5<\lambda\leq1$\verb" Then "\\
\hspace*{\espacio em}\verb"         Evaluate "$\hat{\sigma}_c$\verb" and "$\hat{\tau}_c(\psi(\theta,\chi),\lambda)$\verb"  [Eq."\eqref{interfacelaw}\verb"]"\\
\hspace*{\espacio em}\verb"         "$\hat{t}_c(\theta,\chi,\lambda) = \sqrt{\hat{\sigma}^2_{c}(\psi(\theta,\chi),\lambda)+\hat{\tau}^2_{c}(\psi(\theta,\chi),\lambda)}$\\
\hspace*{\espacio em}\verb"         "$\displaystyle S(\theta,\chi)=\frac{\hat{t}_c(\theta,\chi,\lambda)}{\hat{t}(\theta,\chi)}$\verb" [The critical load factor for "$\theta$\verb" ]"\\
\hspace*{\espacio em}\verb"     Else"\\
\hspace*{\espacio em}\verb"         "$S(\theta,\chi)=\infty$ \verb" [Debond is not allowed at "$\theta$\verb"]"\\
\hspace*{\espacio em}\verb"     Endif"\\
\hspace*{\espacio em}\verb"   Endfor" \\
\hspace*{\espacio em}\verb"   "$S^\infty_c(\chi)=\min\limits_{\theta}S(\theta,\chi)$\verb" and "$\theta_o(\chi)=\arg\!\min\limits_{\!\!\!\!\!\!\!\!\!\!\!\!\theta}S(\theta,\chi)$\\
\hspace*{\espacio em}\verb"   "$\phi^\infty(\chi)=\frac{\pi}{2}\left(\chi-\frac12\right)$\\
\hspace*{\espacio em}\verb"   "$\displaystyle\frac{\sigma_{cx}^{\infty}}{\bar{\sigma}_c}(\chi)=S^\infty_c(\chi)\cos\phi^\infty(\chi)$\verb" and "
$\displaystyle\frac{\sigma_{cy}^{\infty}}{\bar{\sigma}_c}(\chi)=S^\infty_c(\chi)\sin\phi^\infty(\chi)$\\
\hspace*{\espacio em}\verb"Endfor"
\caption{Procedure  for the evaluation of the normalized failure curve $\left({\sigma_{cx}^{\infty}}(\chi)/{\bar{\sigma}_c},{\sigma_{cy}^{\infty}}(\chi)/{\bar{\sigma}_c}\right)$  and  angle  $\theta_o(\chi)$ where the debond initiates, for a circular inclusion subjected to remote biaxial loads.}
\label{procedure}
\end{figure}

The procedure in Fig.~\ref{procedure} predicts the  critical biaxial load for each given load biaxiality parameter $\chi$ leading to the   failure of the first interface point. However, it may be  not clear if this initial infinitesimal debond will further  grows unstably under the same critical load or an additional increase of this load is required to keep the infinitesimal debond growing. This question will be answered applying a numerical procedure like that presented in the next section.

\subsection{Numerical procedure applied to analyse the fibre-matrix debond onset and propagation}\label{NumericalProcedure}
%

 The present    non-linear problem of the  crack onset and propagation along the fibre-matrix interface governed by the  LEBIM is solved by means of the BEM, which is   very suitable for solving this kind of problems where all nonlinearities are placed on the boundaries of the subdomains.  Implementation details  of the  collocational 2D  BEM code   employed and  an overall description of the solution algorithm  can be found in \cite{TavaraEABE2011,TavaraCMES2010,Tavaratesis,graciani2005}. This  algorithm  uses an incremental formulation and a very efficient solution procedure, usually referred to as  sequentially linear analysis, appropriate for the present non-linear problem. The  present BEM model represents a cylindrical inclusion with a radius $a=$7.5 $\mu$m  inside a
relatively large square matrix with side $2\ell=1$~mm. BEM mesh  has 1472 continuous linear boundary elements: two uniform meshes of 720 elements
 discretizing both sides of the fibre-matrix interface (therefore, the polar angle of each element is 0.5$^\circ$) and
32 elements for the external boundary of the matrix, where the remote stresses $\sigma_x^\infty$ and $\sigma_y^\infty$ are applied.   Rigid body motions are removed by the Method F2 introduced in \cite{blazquez1996}, see also \cite{graciani2005}. The  inclusion is  considered initially as bonded to the matrix along its perimeter  by means of a continuous distribution of springs  governed by the LEBIM. The debond onset and propagation is modeled by progressively breaking springs between boundary element nodes placed at both sides of the interface. Thus, the numerical procedure used is driven by the interface crack length and is able to analyse both snap-through and snap-back instabilities of a crack growth.

\section{Results for the fibre-matrix debond onset and propagation}\label{SectionResults}

The aim of this section is to study  the influence of   the  governing  parameters  $\xi$~\eqref{knkt}, $\lambda$~\eqref{interfacelaw}, $\chi$~\eqref{chidef} and $\gamma$~\eqref{gammadef} of the present model on the debond onset and propagation  in the case of the glass-fibre and epoxy-matrix composite (Section \ref{Section_Problem}). Specifically, first, the debond onset is studied focusing in the angle of  debond onset as a function of the remote stress biaxiality (Section~\ref{subsection_Position_crack_onset}) and  by  evaluating  the failure curves in the plane of normalized remote stresses (Section~\ref{subsection_Failure_curves}). Then, debond growth is studied by evaluating load-debond opening curves and load-debond length curves (Section~\ref{subsection_Effect_load_biaxiality}). Finally, an instability analysis of the debond onset and growth is introduced (Section~\ref{subsection_Instability_analysis}). Both analytical and numerical procedures developed are applied wherever feasible, and their results are compared, which allows us to mutually verify the correctness of the  formulation and implementation of these procedures. The analytical procedure is  very suitable for some of the parametric studies presented, nevertheless its range of application is limited to the debond onset characterization  in the present problem of a single fibre  embedded in an infinite matrix.
The scope of the  numerical procedure developed is much wider and it will  allow us to solve complex realistic problems of concurrent debond onset and propagation in    dense fibre packing including random distribution of many fibres with different initial and  boundary conditions (including contact conditions) in future, cf.~\cite{Tavaratesis,TavaraBETEQ2013}.

\subsection{Position of the crack onset}\label{subsection_Position_crack_onset}
\begin{figure}[!htb]
\centering
\subfigure[]{\includegraphics[angle=270,scale=0.47]{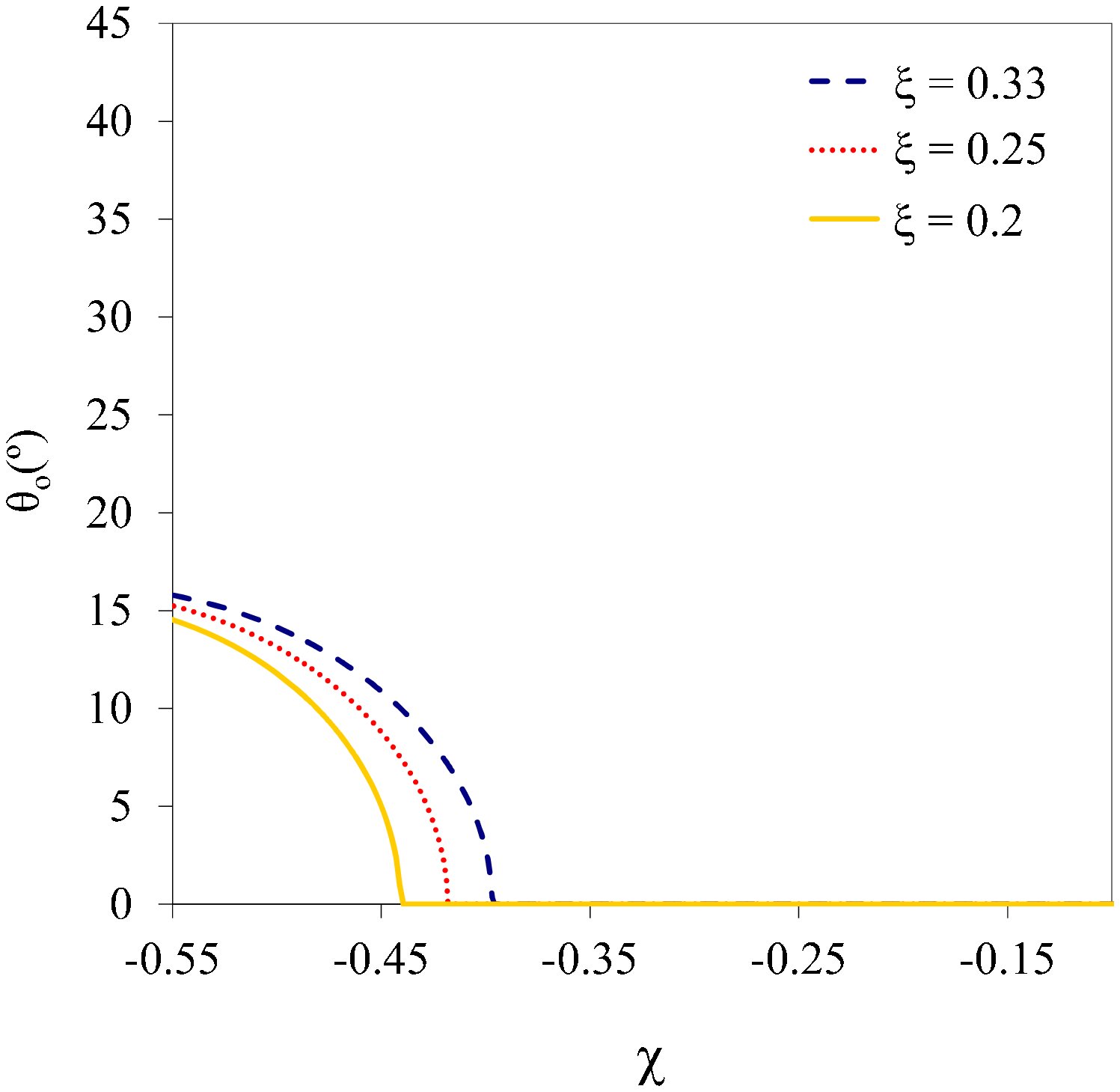}}
\subfigure[]{\includegraphics[angle=270,scale=0.47]{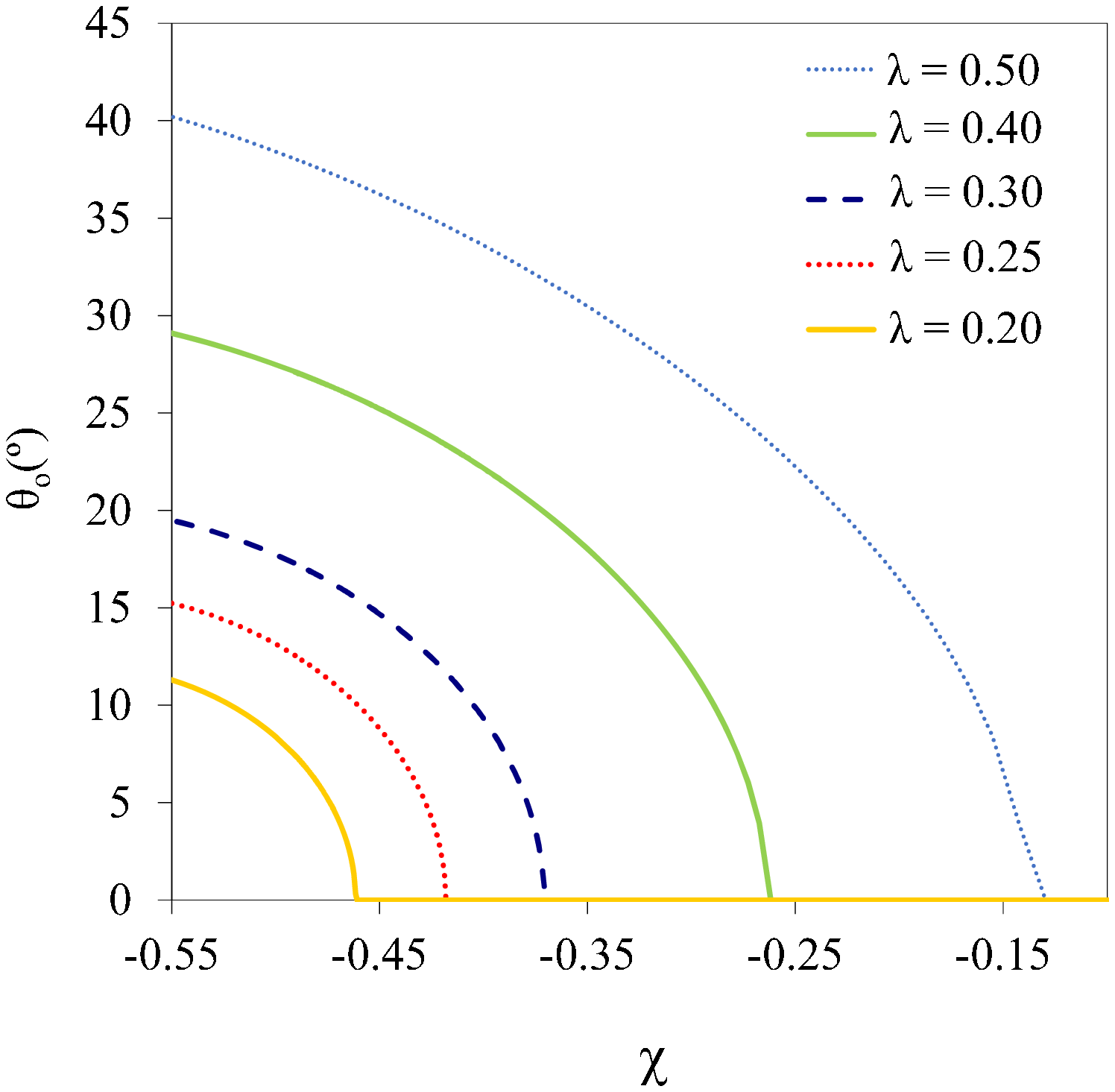}}
\subfigure[]{\includegraphics[angle=270,scale=0.47]{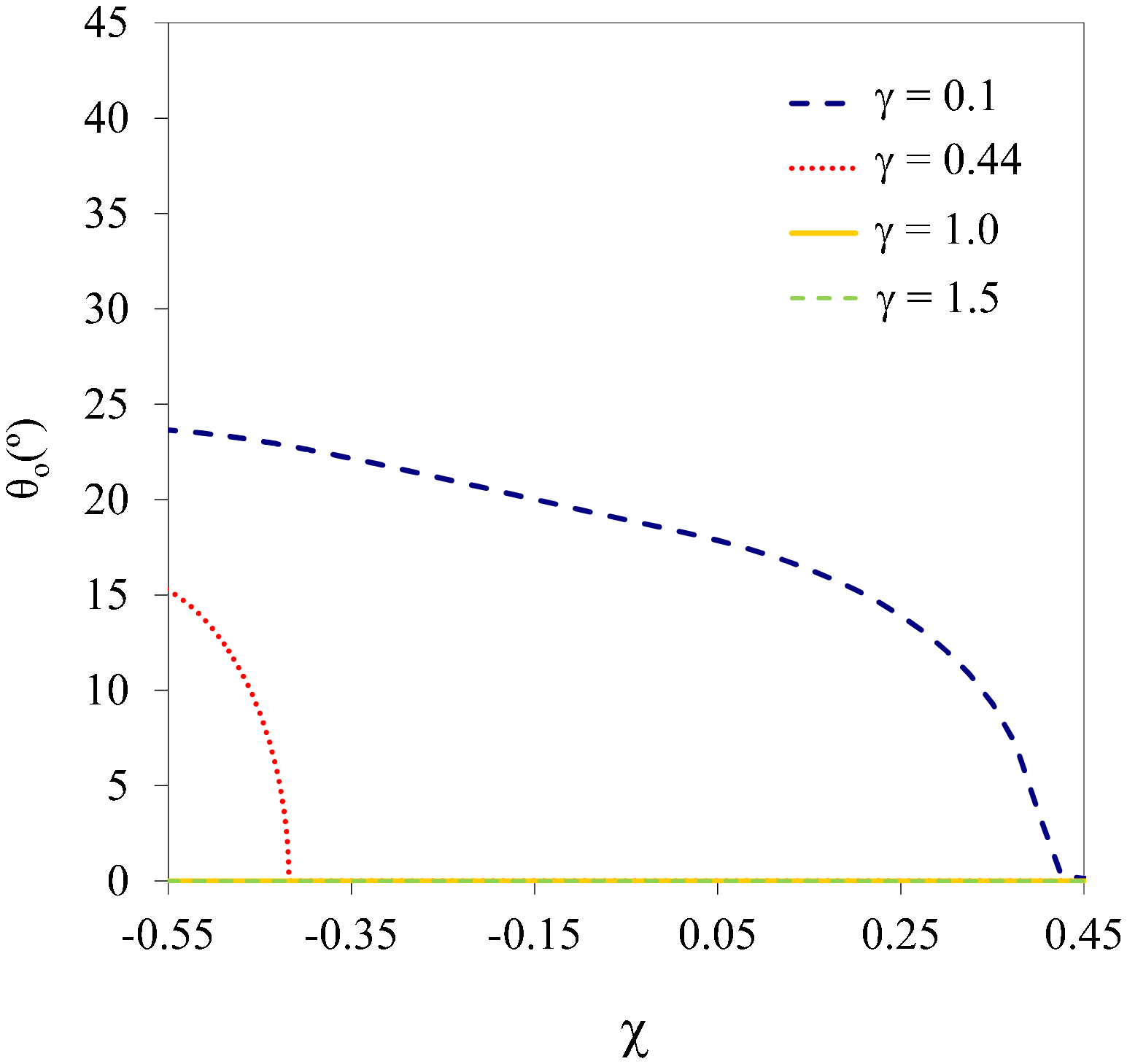}}
\caption{Analytic results. Plots of the crack onset angle $\theta_o$ as a function of the load biaxiality parameter $\chi$   for several values of the  material and structural  parameters (a) $\xi$, (b) $\lambda$ and (c) $\gamma$, and  the following default values where corresponds: $\xi$=0.25,  $\lambda$=0.25,  $\gamma$=0.44.}
\label{thetaochi}
\end{figure}

%
The position where the crack onset occurs,   defined by the angle $\theta_o$, Fig.~\ref{fibmat}, is studied by means
of the  analytic  procedure introduced in Section~\ref{SectionAnalyticProcedure}. Plots of $\theta_{o}(\chi;\xi,\lambda,\gamma)$ in Fig.~\ref{thetaochi} show the influence of different governing parameters on this angle. Notice that for $\chi=1$ (remote equibiaxial tension) all interface points are equivalent and $\theta_o$ is undetermined.

According to these plots of  $\theta_{o}(\chi)$, a bifurcation  takes place  at a particular value of  $\chi$ referred to as bifurcation value $\chi_{b}(\xi,\lambda,\gamma)$. For $\chi_{b}\leq\chi< 1$ the first interface point breaks at  $\theta_o=0^\circ$ in pure fracture  mode I. This behaviour could be expected for   tension dominated remote loads roughly characterized by  $\chi>0$,
 taking into account the distribution of interface tractions \eqref{Gaosolution}
and the failure criterion in Fig.~\ref{sigmataulam}. Nevertheless, as will be seen, there is an exception observed for very brittle configurations.
For  $\chi<\chi_{b}$, a kind of bifurcation is observed due to a sudden variation of $\theta_o$ for $\chi$ below, and close, to $\chi_{b}$. In this case, the interface breaks in a mixed    mode.

 The influence of $\xi$ on $\theta_o$ is depicted  in Fig.~\ref{thetaochi}(a), showing that with increasing value of $\xi$ the bifurcation value $\chi_{b}$  increases slightly as well. Nevertheless, it seems that for decreasing $\chi<-0.5$ all curves tend to a similar value of $\theta_o$.

Fig.~\ref{thetaochi}(b) presents the influence of $\lambda$ on $\theta_o$, showing that with increasing value of $\lambda$ the bifurcation value $\chi_{b}$   increases as well. Thus, for large values of   $\lambda$ a non-symmetric debond initiation is predicted for biaxial tension-compression loading, with tension being only a little lower than compression. For $\chi<\chi_{b}$ the value of $\theta_{o}$ increases with increasing value  of $\lambda$, which could be expected, as the interface failure criterion becomes more sensitive to  the interface shear traction value according to Fig.~\ref{sigmataulam}.

From Fig.~\ref{thetaochi}(c), showing the influence of $\gamma$ on $\theta_o$,  it can be observed that for higher values of $\gamma$  no bifurcation takes place and $\theta_o=0^\circ$, predicting the debond onset in mode I, for  the considered  values of $\chi$, $-0.55\leq\chi<1$. However, for lower values of $\gamma$ a non-symmetric debond is predicted for biaxial tension-compression loading  even for relatively small values of compression load. Actually,  it can be shown    that  for a low value of $\gamma$ and a high  value of $\lambda$, e.g. $\gamma=0.1$ and $\lambda=0.5$,   a non-symmetric debond initiation   would be predicted even for the uniaxial tension. This somewhat surprising behaviour can be explained by the observation that the ratio of the maximum values of $\tau$ to $\sigma$ in \eqref{Gaosolution} is increasing for decreasing $\gamma$ (and/or decreasing  $\chi$) making easier the debond onset in mixed mode. It is remarkable that a similar behaviour  for the uniaxial tension has also been observed in predictions by other models as CZM and FFM in \cite{paggigarcia2013}.

\subsection{Failure curves}\label{subsection_Failure_curves}

\begin{figure}[!ht]
\centering
\subfigure[]{\includegraphics[angle=270,scale=0.4]{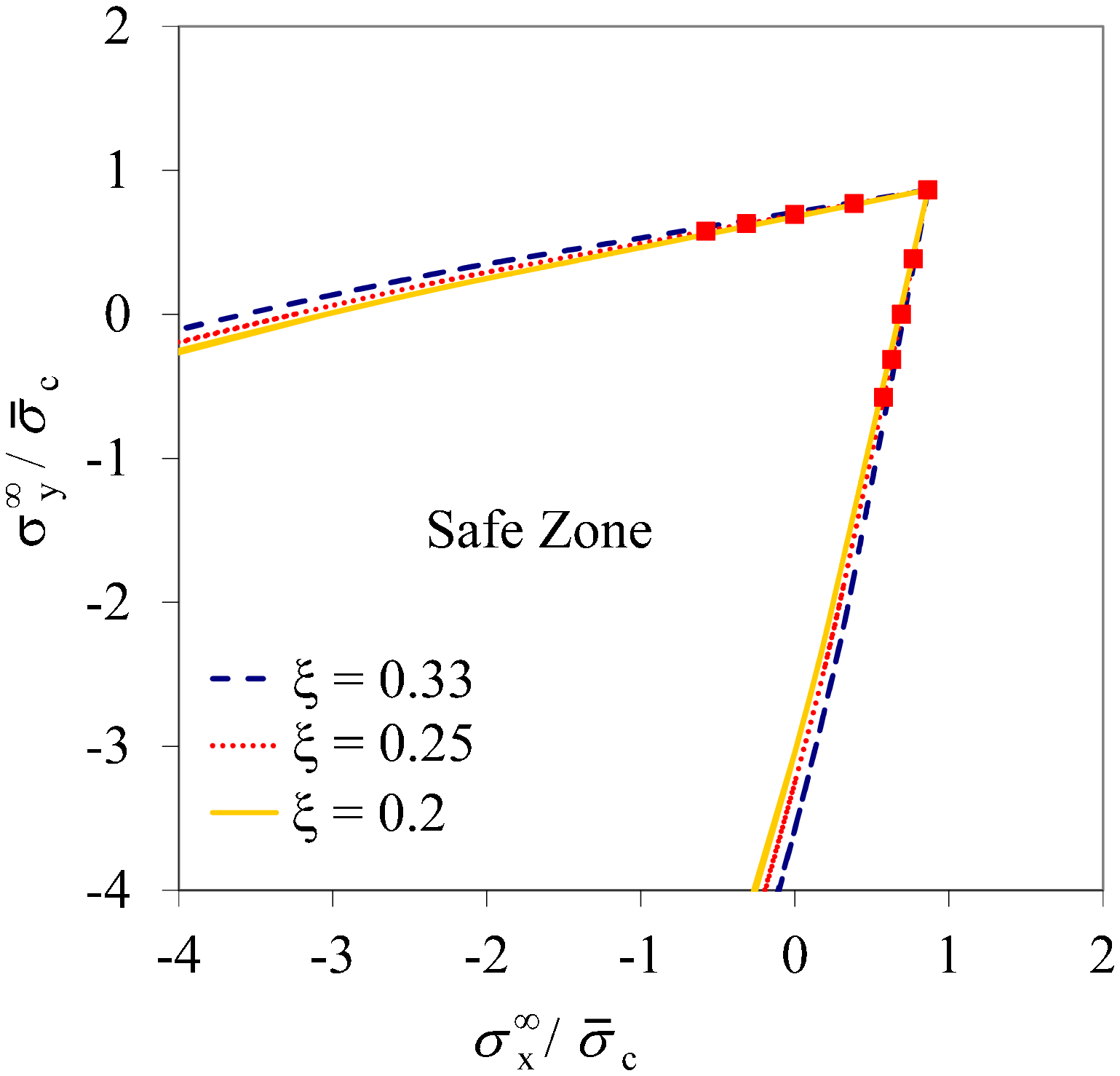}}
%
\subfigure[]{\includegraphics[angle=270,scale=0.4]{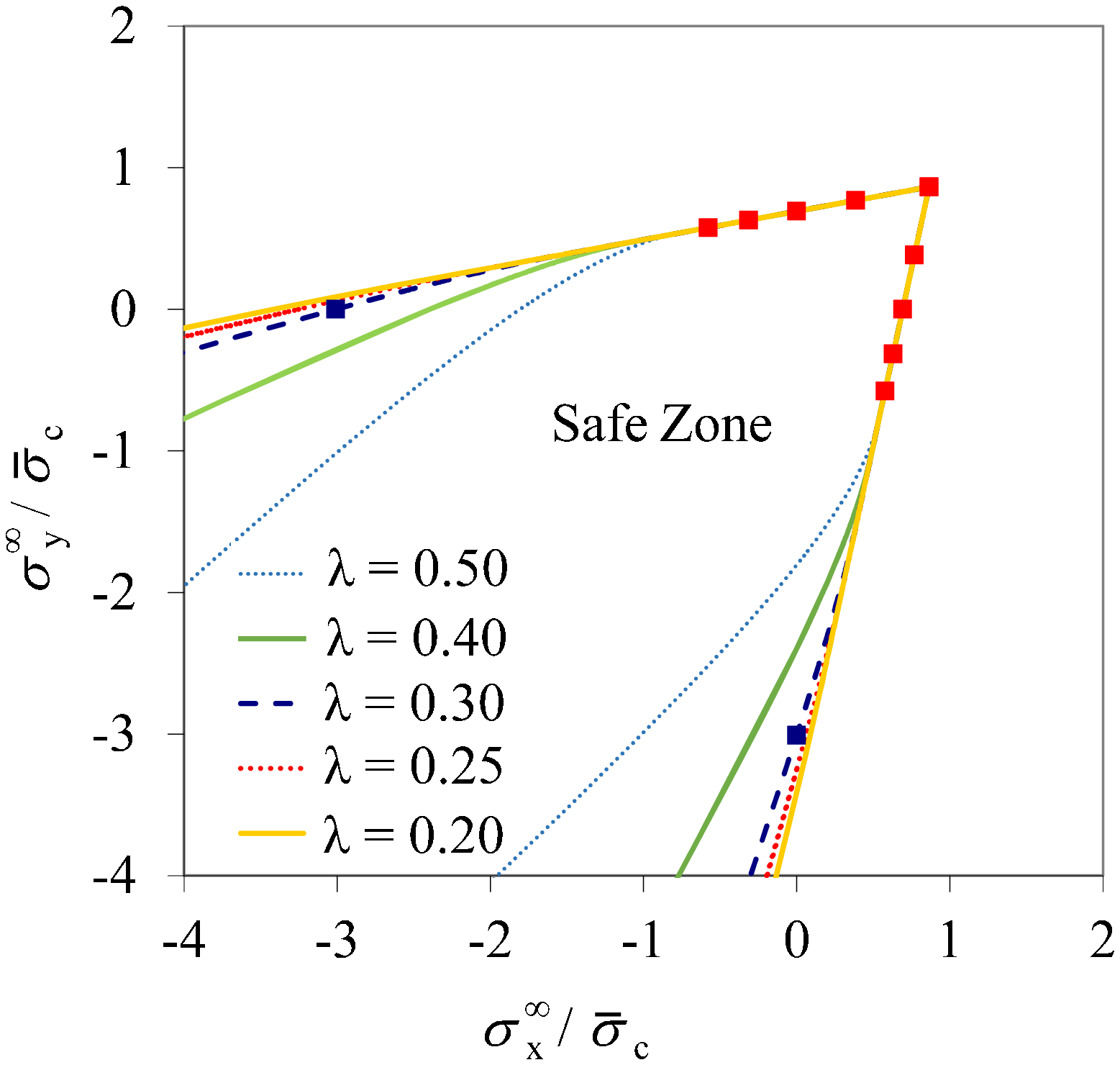}}
%
\subfigure[]{\includegraphics[angle=270,scale=0.4]{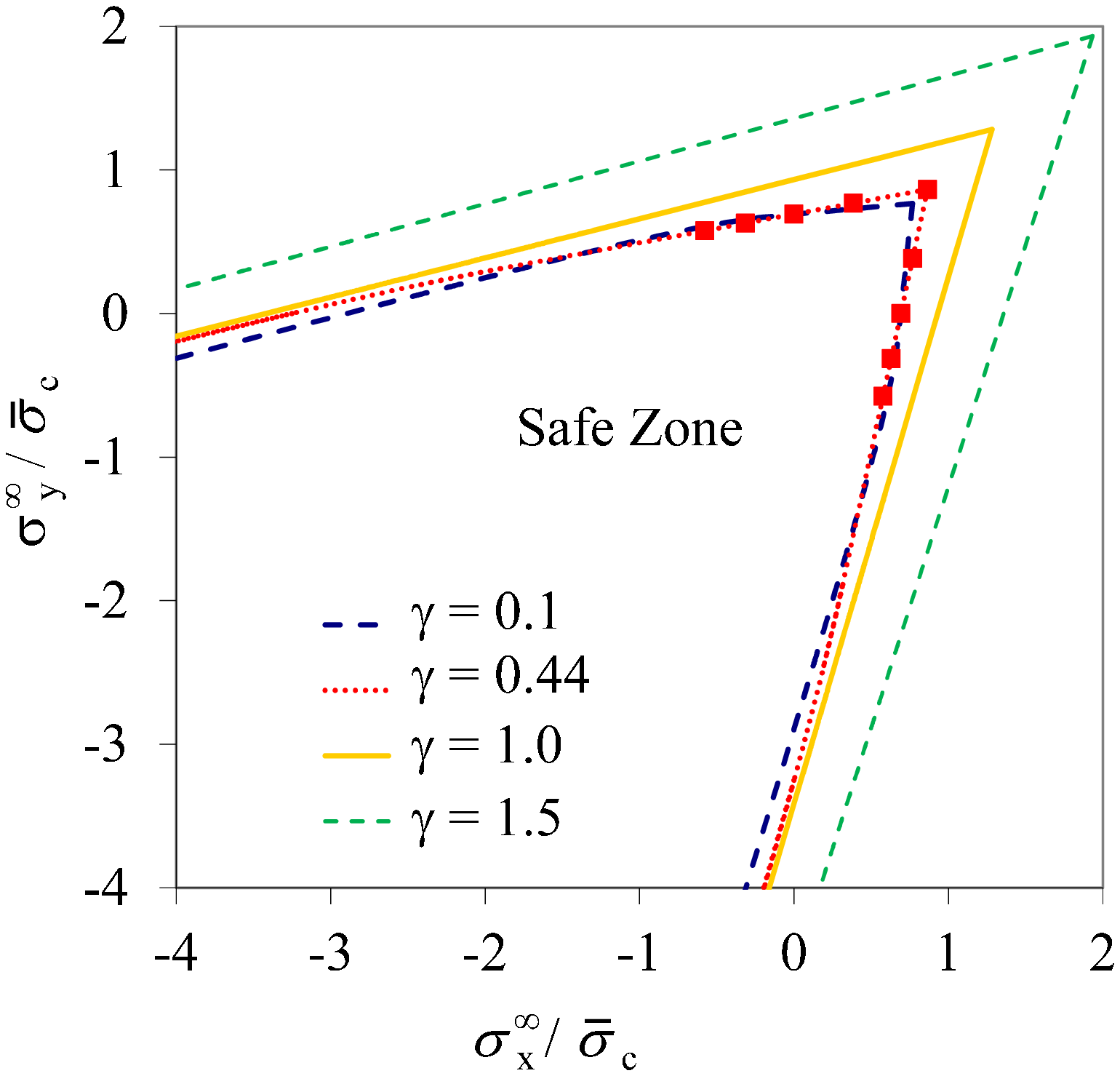}}
\caption{Analytic (curves) and numerical (marks) results. Normalized failure curves of a circular inclusion under biaxial transversal loads for several values of the  material and structural  parameters (a) $\xi$, (b) $\lambda$ and (c) $\gamma$, and  the following default values where corresponds: $\xi$=0.25,  $\lambda$=0.25,  $\gamma$=0.44. }
\label{biaxcri}
\end{figure}

%
Fig.~\ref{biaxcri} presents  failure curves parameterized by the load biaxiality parameter $\chi$ and representing the normalized critical remote stresses leading to the breakage of the first point (spring) of an initially undamaged inclusion-matrix interface. Analytical and numerical results are represented by continuous lines and  marks, respectively.
These plots show the influence of the   material ($\xi$ and $\lambda$) and structural ($\gamma$) dimensionless parameters of the problem on the failure curve shape and location.

Regarding   the influence of the load biaxiality parameter $\chi$, it is easy to observe in all the plots in Fig.~\ref{biaxcri} that, considering $\sigma_{x}^\infty\geq\sigma_{y}^\infty$ (thus looking at the right-bottom branch of failure curves),   for decreasing values of $\chi$ the critical remote stress $\sigma_{cx}^\infty$ decreases quite significantly. In particular, a relevant compression $\sigma_{y}^\infty<0$ makes a debond onset easier, $\sigma_{cx}^\infty$ being significantly smaller than in the case    of a   tension $\sigma_{y}^\infty>0$ or even in the case of the uniaxial tension ($\chi=0.5$) when $\sigma_{y}^\infty=0$.

The rather weak influence of the ratio of the interface stiffnesses $\xi$ on the fibre-matrix failure curve can be observed in Fig.~\ref{biaxcri}(a) obtained
by varying $\xi$ ($\xi$=0.20, 0.25 and 0.33) and  keeping constant the fracture mode-sensitivity parameter $\lambda=0.25$ and the brittleness number $\gamma$=0.44.
For lower values of $\xi$ the critical loads are only slightly lower, this influence being mostly visible for $\chi<0$, and in particular for the case of the uniaxial compression ($\chi=-0.5$).

The influence of the fracture mode-sensitivity parameter $\lambda$ on the  failure curve is studied in Fig.~\ref{biaxcri}(b),
by varying $\lambda$ ($\lambda$=0.2, 0.25, 0.3, 0.4 and 0.5) and   keeping constant $\xi=0.25$ and $\gamma=0.44$.
 There is no influence of $\lambda$ on the failure curve for  $-0.12  \lesssim \chi \leq 1$,
  because in this range the crack onset occurs  at $\theta_o=0^\circ$, see Fig.~\ref{thetaochi}(b), which, in view of the symmetry of the stress solution \eqref{Gaosolution}, means that shear tractions vanish there, and  consequently the interface  breaks in mode I at this point.
Nevertheless, for larger values of compressions $\sigma_{y}^{\infty}$, i.e. $\chi \lesssim -0.12$,
  the crack onset   changes its position  given by $\theta_o > 0^\circ$, see Fig.~\ref{thetaochi}(b),
   the interface  the breaking in a mixed mode there. This leads to a strong   influence of $\lambda$ on the shape of failure curves for this range of $\chi$, the critical loads being significantly lower  for larger values of $\lambda$, because the interface strength strongly decreases with increasing $\lambda$ according to \eqref{interfacelaw} and Fig.~\ref{sigmataulam}.

The influence of the brittleness number  $\gamma$ on the  failure curve is shown in Fig.~\ref{biaxcri}(c),
by varying $\gamma$  ($\gamma$=0.1, 0.44,  1 and 1.5) and   keeping constant $\xi=0.25$ and $\lambda=0.25$. While the variations of failure curves for small values of $\gamma$ (brittle configurations) predicting  small critical loads,  are hardly visible,   a quite strong influence  of $\gamma$ on the position of failure curves  is observed  for  larger values of $\gamma$ (tough configurations) predicting large  critical   loads. Notice that, in view of the dependence of $\gamma$ on the inclusion radius $a$ \eqref{gammadef}, the variations of the failure curves with $\gamma$ represent in fact a size effect of $a$ on the crack onset, cf.~\cite{mantic2009,TavaraEABE2011,manticgarcia2012,Paggi2005}.

As can be observed from  Fig.~\ref{biaxcri}, an excellent agreement is achieved between the analytical and numerical procedures for several tension dominated biaxial loads (with parameters $\xi=0.25$, $\lambda=0.25$ and $\gamma=0.44$)  and a  uniaxial compression load ($\xi=0.25$, $\lambda=0.3$ and $\gamma=0.44$). Recall that   the present formulation of the LEBIM, see Fig.~\ref{sigmataulam}, allows   studying also crack growth under compressions in presence of large shear tractions at the crack tip which are typically  associated to contact between the crack faces in a zone adjacent to this tip. This capability   allows us to model crack onset and growth even in  the  case of remote compressions   applied in both directions, i.e. for $\chi<-0.5$.

As mentioned above, one of the reasons of larger differences between some failure curves shown in Fig.~\ref{biaxcri} are  the   variations of the  crack onset position given by the angle $\theta_o$.

\subsection{Effect of  the load biaxiality on the fibre-matrix debond onset and growth}\label{subsection_Effect_load_biaxiality}
The effect of the load biaxiality on the debond onset and growth is  studied by the numerical procedure presented in Section \ref{NumericalProcedure}. It will be shown  that the failure curves presented in Fig.~\ref{biaxcri}, referring to the breakage of the first interface point, actually represent the initiation of an unstable crack growth along the inclusion-matrix interface. The  default values   $\xi$=0.25,   $\lambda$=0.25 and   $\gamma$=0.44 are chosen for the following numerical study.

\begin{figure}[!htb]
\centering
\subfigure[]{\includegraphics[angle=270,scale=0.47]{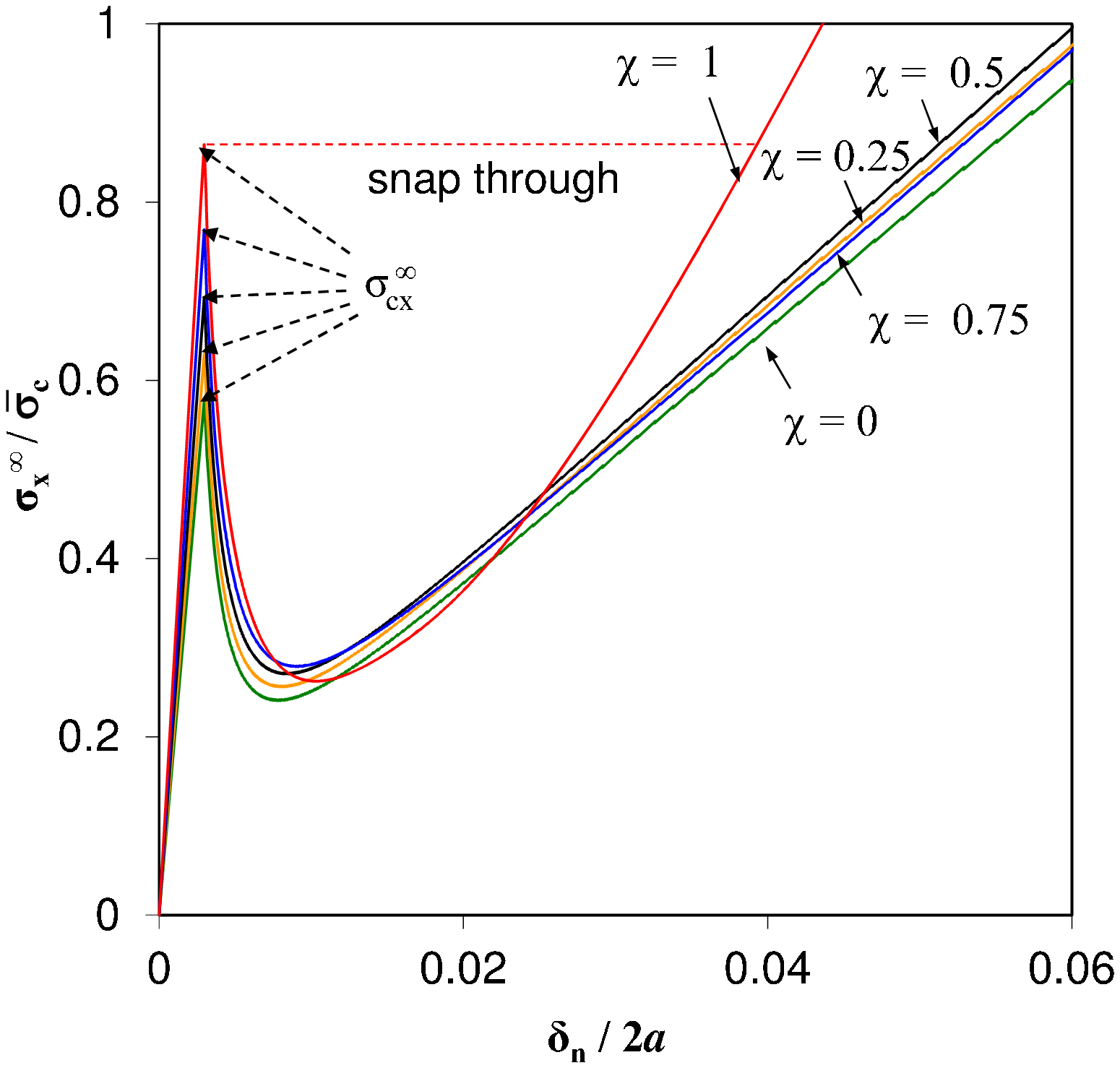}}\hspace{0.2cm}
\subfigure[]{\includegraphics[angle=270,scale=0.47]{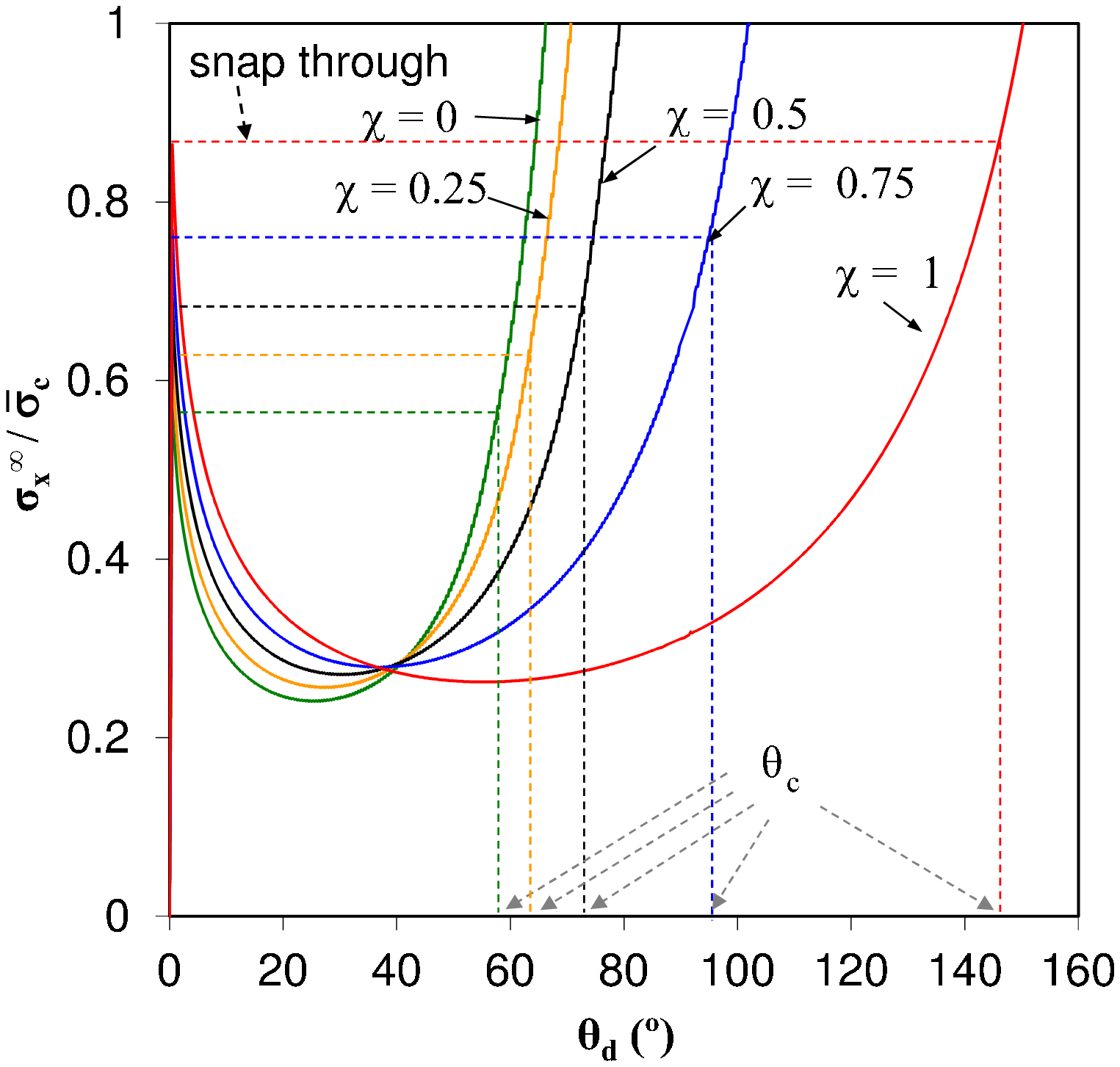}}
\caption{Numerical results. (a) The normalized applied stress with respect to the normal relative displacements $\delta_n$ at point A, see Fig.~\ref{fibmat}, and (b) The normalized  applied stress with respect to the semidebond angle $\theta_d$ for different biaxial loads combinations, with $\xi=0.25$, $\lambda$=0.25 and  $\gamma$=0.44.}
\label{sigmadth}
\end{figure}

In Fig.~\ref{sigmadth} and Table \ref{restab1} the numerical results obtained for different values of the  load  biaxiality parameter   $\chi=0,0.25, 0.5, 0.75$ and $1$
are presented. Recall that $\chi=0.5$ corresponds to the case of uniaxial tension in the $x$-direction ($\sigma_y^{\infty}=0$).
%

In Fig.~\ref{sigmadth}(a), the normalized  remote stress $\sigma_x^{\infty}/\bar{\sigma}_{c}$  is
plotted as a function of the normal relative displacement (opening), $\delta_n$,
evaluated at the point A$(a,\theta_o=0^\circ)$   defined in Fig.~\ref{fibmat}(b). The (minimum) remote
stress value  that is needed to initiate crack growth (in simple terms, the stress
that is needed to break the first spring  in the present
discrete model of the  interface) is called critical stress, $\sigma_{cx}^{\infty}$, and corresponds
to the local maximum of a function  shown in Fig.~\ref{sigmadth}(a).
It can  also be observed in Fig.~\ref{sigmadth}(a) that after reaching the critical
stress, $\sigma_{cx}^{\infty}$, the crack growth becomes unstable, requiring
smaller values of the remote tension to cause further crack growth.
Thus, an instability phenomenon  called snap-through is predicted   in the case of external load or displacement control, see Section \ref{subsection_Instability_analysis}.

The variations of   the local maxima values in Fig.~\ref{sigmadth}(a) confirm the conclusion observed previously in Fig.~\ref{biaxcri} that the critical stress $\sigma_{cx}^\infty$ decreases with decreasing $\chi$, see also  Table \ref{restab1}.

In   Fig.~\ref{sigmadth}(b),  the normalized  remote stress $\sigma_{x}^{\infty}/\bar{\sigma}_{c}$
is plotted versus the semidebond angle $\theta_d$
defined in Fig.~\ref{fibmat}(b).
 An estimation of the critical semidebond angle $\theta_c$ defined as the
semidebond angle $\theta_d$ reached at the end of the initial unstable crack growth, keeping  the remote stress  $\sigma_{cx}^{\infty}$ constant, is also indicated in this figure. In general, $\theta_c$ increases with increasing $\chi$ in the range studied, see also Table \ref{restab1}.
When $\mu\gtrsim 0.75$, i.e. when significant remote tensions are applied in both axes,
 $\theta_c>90^\circ$. Thus, an unstable debond growth     is predicted  along a very large portion of the fibre-matrix interface.

\renewcommand{\multirowsetup}{\centering}
\begin{table}[!htb]
\caption{The normalized critical stress for crack onset  $\sigma_{cx}^{\infty}$  and critical semidebond angle  $\theta_c$, for different  values of $\chi$, with $\xi=0.25$, $\lambda$=0.25 and  $\gamma$=0.44.}
\begin{center}
\begin{tabular}{c|c|c|c|c|c|}
\cline{2-6}
&\multicolumn{5}{|c|}{$\chi$} \\
\cline{2-6}
&  0   &   0.25 &   0.5   &  0.75  &   1   \\
\cline{1-6}
\multicolumn{1}{|c|}{$\sigma_{cx}^{\infty}/\bar{\sigma}_c$} & 0.573 & 0.629 & 0.692 & 0.769 & 0.864 \\
\multicolumn{1}{|c|}{$\theta_c$ ($^\circ$) }                & 58.25 & 63.25 & 72.75 & 95.25 & 146.0 \\
\hline
\end{tabular}
\end{center}
\label{restab1}
\end{table}

Actually, the prediction of an unstable crack growth up to the critical semiangle $\theta_c$ is the key result obtained by the numerical  solution of the present problem,  as the values of    $\sigma_{cx}^{\infty}$  and $\theta_o$ can also be obtained by the analytical procedure presented. An excellent agreement    between the analytic and  numerical results is remarkable.

\subsection{Instability analysis of the fibre-matrix debond onset and growth}\label{subsection_Instability_analysis}

\begin{figure}[!htb]
\centering
\subfigure[]{\includegraphics[angle=270,scale=0.47]{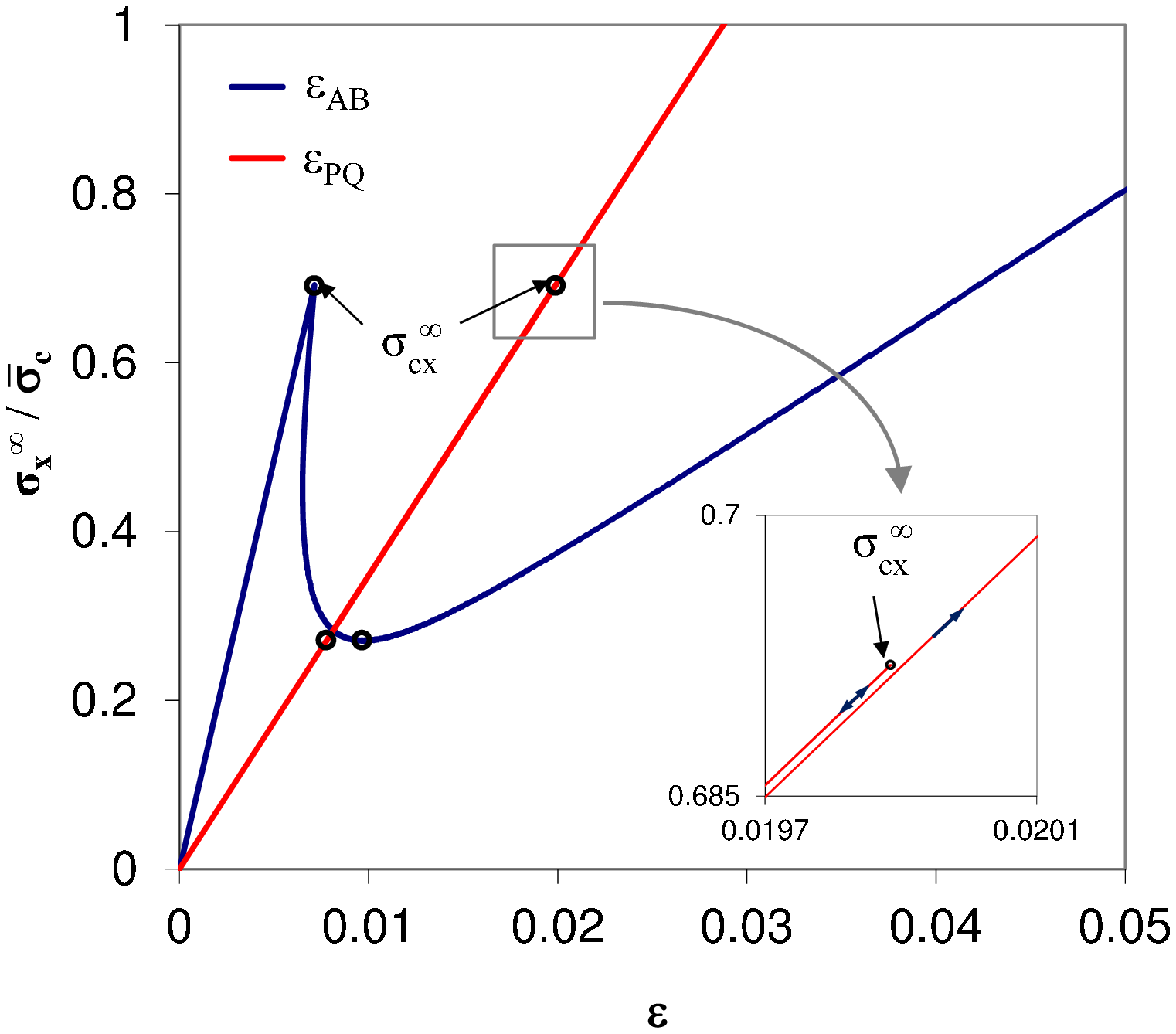}}\hspace{0.2cm}
\subfigure[]{\includegraphics[angle=270,scale=0.47]{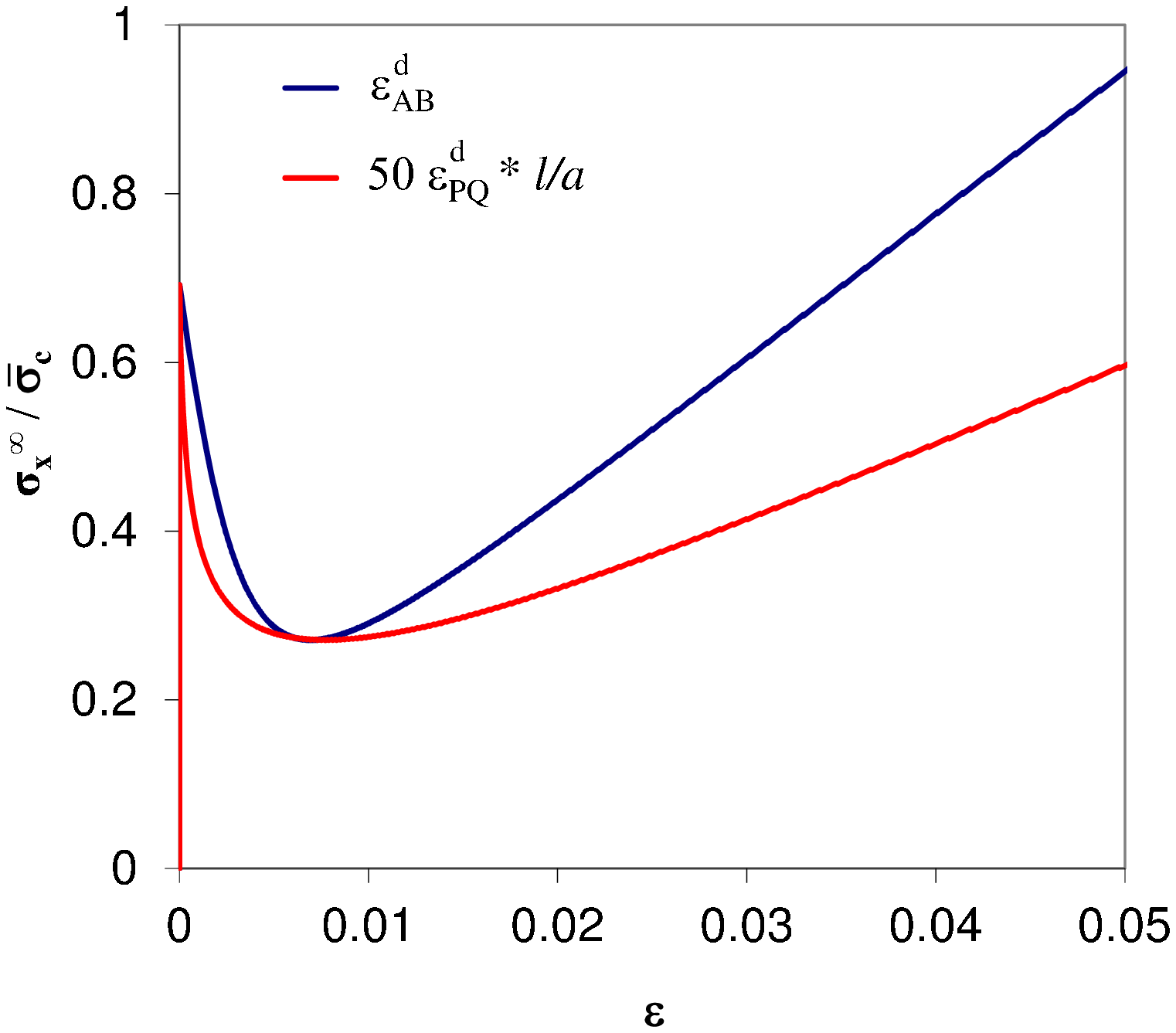}}
\caption{Numerical  results. The normalized  applied stress $\sigma_{x}^\infty$ with respect to (a) the averaged longitudinal strains, $\varepsilon_{AB}=\varepsilon_{AB}^{e}+\varepsilon_{AB}^{d}$ and $\varepsilon_{PQ}=\varepsilon_{PQ}^{e}+\varepsilon_{PQ}^{d}$, and (b) the additional averaged longitudinal strain due to debond, $\varepsilon_{AB}^{d}$ and the scaled one $50\frac{\ell}{a}\varepsilon_{PQ}^{d}$, with $\chi=0.5$, $\xi=0.25$, $\lambda$=0.25 and  $\gamma$=0.44.}
\label{InstabilityPlot}
\end{figure}

In the following section, the instability behaviour (snap-through) observed in Fig.~\ref{sigmadth} will be analysed in order to check why it may appear in an external load or displacement control. Although only the case of uniaxial tension $(\chi=0.5)$, with the default values of $\xi$, $\lambda$ and $\gamma$, is considered for the sake of brevity, the results would be similar for other values of the governing dimensionless parameters. Fig.~\ref{InstabilityPlot}(a) shows  the normalized applied remote stress $\sigma_{x}^\infty/\bar{\sigma}_{c}$ versus the averaged longitudinal strain $\varepsilon$ along the segments, $AB$ and $PQ$,  between two pairs of  points of the matrix placed on the $x$-axis and symmetrically situated with respect to the origin. The coordinates of the end points of $AB$ are $(x=\pm a,y=0)$ and of $PQ$   $(x=\pm \ell,y=0)$, where $a$ is the fibre radius and $\ell$ the half-length of the matrix square cell side, $\ell/a=66.7$ in the present study. $\varepsilon^{e}$ represents the averaged longitudinal strain for a purely linearly elastic fibre-matrix interface with no debond, while  $\varepsilon^{d}$ is the  additional averaged longitudinal strain   due to debond $(\varepsilon^{d}=\varepsilon-\varepsilon^{e})$.   For a similar additive decomposition of relative displacements, see~\cite{BazantCedolin1991} (Ch. 12 therein).

The diagrams $\sigma_{x}^\infty-\varepsilon$ in both cases (considering segments $AB$ and $PQ$)  exhibit    cusp snapback instability~\cite{carpinteri1989a} after the peak point (bifurcation point)  where the debond onset occurs. Actually, this kind of instability    also appears for all intermediate  segments between    $AB$ and $PQ$. While the snapback instability is easily observable in the curve $\sigma_{x}^\infty-\varepsilon_{AB}$   in Fig.~\ref{InstabilityPlot}(a), this instability is not visible by naked-eye  in the curve $\sigma_{x}^\infty-\varepsilon_{PQ}$, as the  curve branches before and after the peak point are extremely close to each other, visually coinciding in the plot, because  the matrix cell is very large with respect to the fibre. As the effect of the  debond onset and growth on the fibre-matrix interface is hardly visible on this plot,   a   zoomed view of this curve with its cusp is also included in Fig.~\ref{InstabilityPlot}(a) to show this  instability behaviour. Obviously the values of $\sigma_{x}^\infty/\bar{\sigma}_{c}$   at the local maxima (peak point) and minima  in both curves coincide (values 0.692 and 0.2704, respectively)  as   indicated in the curve plots. It means   that after the debond onset,  we may  decrease  the applied load significantly, up to 39\% of the critical load in the peak, keeping   a continuous propagation of the debond.

Moreover, to understand better the post-peak behaviour, diagrams   $\sigma_{x}^\infty-\varepsilon^{d}$ are plotted  in Fig.~\ref{InstabilityPlot}(b). The value  of  $\varepsilon_{PQ}^{d}$, which strongly depends on the cell size $\ell$ (as a consequence of the Saint-Venant and superposition principles),  is scaled by an arbitrary factor $50\, \frac{\ell}{a}$ resulting in a value  very similar to that of  $\varepsilon_{AB}^{d}$. The initial very steep negative  slope of these diagrams indicates  that    according to the   present model, using the LEBIM of the fibre-matrix interface, the debond onset and growth exhibits cusp snapback  instability typical for a brittle structural behaviour.  This observation is quite different from a smooth snapback instability observed in some cases in~\cite{Paggi2005,paggigarcia2013} using a CZM of the fibre-matrix interface.

Summarizing the above analysis,  the  curve $\sigma_{x}^\infty-\varepsilon_{PQ}$ shows that under both  load and displacement control at the outer boundaries of the matrix cell a sudden and large breakage of the fibre-matrix interface is predicted by the present model. Notice that, the debond onset and growth could develop, at least hypothetically,  in a stable manner if it would be  controlled by the crack opening $\delta_{n}$ according to Fig.~\ref{sigmadth}(a).


\section{Concluding remarks}
 A new linear elastic - (perfectly) brittle interface model (LEBIM)
has been used to characterize the onset and growth of the debond
at a single fibre embedded in an infinite matrix
subjected to biaxial transverse loads $\sigma_{x}^\infty\geq\sigma_{y}^\infty$, Fig.~\ref{fibmat}. Both analytic and numerical procedures have been  devised and exploited to study this problem. The analytic procedure has been used  in the parametric studies regarding debond onset   and for testing the numerical procedure implemented in a collocation BEM code, whereas the numerical procedure is quite general and is  currently applied  to the numerical analysis of debond onset and growth in dense fibre packing representing a portion  of a real unidirectional composite lamina, with several fibres, under biaxial transverse loads~\cite{TavaraBETEQ2013}.

A comprehensive parametric study of this single-fibre debond problem analysing the influence of all the  dimensionless parameters governing the problem:   $\chi$ - load biaxiality \eqref{chidef}, $\xi$ - ratio of the interface shear and normal stiffnesses \eqref{knkt},  $\lambda$ -  sensitivity to interface fracture mode mixity  \eqref{interfacelaw}, and $\gamma$ - brittleness number \eqref{gammadef}, in addition to the elastic properties of fibre and matrix,  has been carried out. To the best knowledge of the authors no similar parametric study has been presented before neither for the LEBIM nor CZMs.

Using a general analytical solution for tractions at the undamaged linear-elastic fibre-matrix interface
 under  uniform far-field biaxial  transverse  stresses  and assuming the LEBIM,
 quite universal failure curves in the plane of  normalized far-field stresses
$(\frac{\sigma_{x}^\infty}{\bar{\sigma}_c},\frac{\sigma_{y}^\infty}{\bar{\sigma}_c})$, where $\bar{\sigma}_c$ is the interface tensile strength, have been generated. These curves, parameterized by  $\chi$,  depend  only on a few dimensionless parameters $\xi$, $\lambda$, $\gamma$, and $E_m/E_f$, $\nu_m$ and $\nu_f$.   In particular, the elastic properties   $E_m$, $E_f$, $\nu_m$ and $\nu_f$ corresponding to a glass-epoxy composite have been considered. It can be observed from  these curves, that with decreasing $\chi$ the critical load  $\sigma_{cx}^\infty$ decreases as well,
i.e. a compression  $\sigma_{y}^\infty$ makes easier
crack onset leading to  a lower critical tension load  $\sigma_{cx}^\infty$, and viceversa a tension $\sigma_{y}^\infty$  difficulties crack onset leading to  a larger value of  $\sigma_{cx}^\infty$. These observations agree with previous experimental results in~\cite{ParisCorrea2003}.

  The debond onset angles $\theta_o(\chi)$ associated to these failure curves have also been evaluated analytically. A bifurcation from the zero value of $\theta_o$, predicting a debond onset in  mixed   mode,  typically occurs for a magnitud of the compression load $\sigma_{y}^\infty$  larger than the tension load  $\sigma_{x}^\infty$, i.e. for $\chi<0$. Nevertheless, in  very brittle configurations characterized by $\gamma \simeq 0$ such a bifurcation can occur for small or vanishing values of $\sigma_{y}^\infty$.


The observed influence of the governing dimensionless parameters on the shape and location of the failure curves and the debond onset angle is summarized in the following: a)   $\xi$   has only a slight influence on the shape and no influence on the position of the failure
curves, also its influence on $\theta_o$ is quite small; b)  $\lambda$ typically has no influence on the debond onset for tension dominated loads as the interface  breaks at $\theta_o$ under pure mode I (except for  very brittle configurations with $\gamma \backsimeq 0$), but it has
a quite relevant influence on $\theta_o$, in particular on its bifurcation point position, for compression dominated loads, consequently $\lambda$ shows  some influence on the shape of failure curves   for such loads, particularly for   $\lambda \simeq 0.5$;
c) $\gamma$ has a strong  influence on   the position of failure curves    for tough  configurations $(\gamma\gtrsim 1)$,  while for brittle configurations its influence on   the position of failure curves  is rather weak    showing, however, some influence on their shape.  The influence of  $\gamma$ on  $\theta_o$ is quite relevant for $\gamma \backsimeq 0$.

From the numerical results obtained, it can be observed that when the remote load  reaches its critical value given by $\sigma_{cx}^{\infty}$,
the subsequent debond growth up to the critical semidebond angle $\theta_c$ is unstable, an instability phenomenon called snap-back taking place.
A parametric study shows that $\theta_c$ increases with increasing $\chi$ in the range studied,  eventually    very large debonds  with $\theta_c > 90^\circ$ are predicted  when similar tensions are applied in both directions.

From the above analytical and numerical results it appears that the new LEBIM formulation introduced   adequately  describes the behavior of the fibre-matrix system, predicting expected behaviour where some experimental results are available~\cite{ParisCorrea2003} and also being in a quite good agrement with other analytical and numerical studies~\cite{ParisCorrea2007,Correa2008a,Correa2008b,manticgarcia2012,Correa2013,paggigarcia2013}.
An important novelty  with respect to the previous LEBIM formulations in~\cite{TavaraEABE2011,TavaraCMES2010} is that the new formulation is able to model
interface crack onset and growth  in presence of   compressive interface tractions, in particular when the crack is closed with crack faces in frictionless   contact.

It has been shown that the present LEBIM implementation in a BEM code is an efficient computational tool for an interface  crack onset and mixed mode crack growth modeling. This tool can be useful not only
 for an analysis of fibre-matrix
debonding under biaxial transverse loads as carried out in the present work and in~\cite{TavaraBETEQ2013}, but also in other problems
 as interlaminar fracture toughness
tests of symmetric and non-symmetric laminates 
and delaminations   in cross-ply laminates.


\section*{Acknowledgements}
The work was supported by the Junta de Andaluc\'{\i}a (Projects of
Excellence TEP-1207, TEP-2045 and TEP-4051), the Spanish
Ministry of Education and Science (Projects TRA2006-08077 and MAT2009-14022) and  Spanish Ministry of Economy and   Competitiveness (Projects  MAT2012-37387 and DPI2012-37187).



\bibliographystyle{model3-num-names}
\bibliography{tavara}

\end{document}

%% file: defsR.tex
\def\beq{\begin{equation}}
\def\eeq{\end{equation}}

\def\XXint#1#2#3{{\setbox0=\hbox{$#1{#2#3}{\int}$ }
\vcenter{\hbox{$#2#3$ }}\kern-.55\wd0}}

\def\epsilon{\varepsilon}

\def\beq{\begin{equation}}
\def\eeq{\end{equation}}
\def\beqa{\begin{eqnarray}}
\def\eeqa{\end{eqnarray}}

